\documentclass[prl,twocolumn]{revtex4-2}
\usepackage{amsmath,amssymb,graphicx,dcolumn,numprint}
\usepackage[usenames,dvipsnames,svgnames,table]{xcolor}
\usepackage[colorlinks=true, urlcolor=Blue, linkcolor=Blue, citecolor=Blue, pdftex]{hyperref}

\npthousandsep{\,}
\npdecimalsign{.}

\def\<{\langle}
\def\>{\rangle}
\newcolumntype{.}{D{.}{.}{6}}
\newcolumntype{q}{D{.}{.}{1}}
\newcolumntype{Q}{D{.}{.}{2}}

\newcommand{\PA}{Physica}
\newcommand{\PAA}{\PA~A}

\newcommand{\JSTAT}{J.~Stat.~Mech.}
\newcommand{\JSP}{J.~Stat.~Phys.}

\newcommand{\PR}{Phys.~Rev.}
\newcommand{\PRL}{Phys.~Rev.~Lett.}

\newcommand{\PRB}{Phys.~Rev. B}

\newcommand{\PRE}{Phys.~Rev. E}

\newcommand{\JPCM}{J.~Phys.: Condens.~Matter}
\newcommand{\JPAOLD}{J.~Phys.~A: Math.~Gen.}

\newcommand{\physrep}{Phys.~Rep.}

\newcommand{\JSF}{J.~Stat.~Phys.}
\newcommand{\PLB}{Phys.~Lett.~B}
\newcommand{\RMP}{Rev.~Mod.~Phys}

\begin{document}

\title{Boundary Critical Behavior of the Three-Dimensional Heisenberg Universality Class}
\author{\firstname{Francesco} \surname{Parisen Toldin}}
\email{francesco.parisentoldin@physik.uni-wuerzburg.de}
\affiliation{\mbox{Institut f\"ur Theoretische Physik und Astrophysik, Universit\"at W\"urzburg, Am Hubland, D-97074 W\"urzburg, Germany}}
\begin{abstract}
  We study the boundary critical behavior of the three-dimensional Heisenberg universality class, in the presence of a bidimensional surface. By means of high-precision Monte Carlo simulations of an improved lattice model, where leading bulk scaling corrections are suppressed, we prove the existence of a special phase transition, with unusual exponents, and of an extraordinary phase with logarithmically decaying correlations. These findings contrast with na\"ive arguments on the bulk-surface phase diagram, and allow us to explain some recent puzzling results on the boundary critical behavior of quantum spin models.
\end{abstract}

\maketitle

{\it Introduction.}---Critical phenomena in the presence of boundaries is a fertile source of interesting phenomena, and has attracted numerous experimental \cite{Dosch-book} and theoretical \cite{Binder-83,Diehl-86,Pleimling-review} investigations. In the simplest setting, one considers a $d$-dimensional system bounded $(d-1)$-dimensional surface, breaking the translation symmetry. For a critical system, the behavior at the surface is remarkably different than the bulk one. In fact, standard renormalization-group (RG) arguments predict that a given bulk universality class (UC) potentially splits into different surface UCs \cite{Diehl-86,Cardy-book}, resulting in a rich bulk-surface phase diagram.
Surface UCs also determine the critical Casimir force \cite{FG-78,Krech-94,Krech-99,Gambassi-09,GD-11,MD-18}.
For classical models,
one generically distinguishes between the surface ordinary UC, where the surface exhibits critical behavior as a consequence of a critical bulk,
the surface critical behavior in the presence of a disordered bulk (when such a transition exists), and the surface extraordinary UC, found for a critical bulk and strong enough surface enhancement. Finally, in the bulk-surface phase diagram these three transition lines meet at a multicritical point, the so-called special UC \cite{Binder-83,Diehl-86}.
In this framework, one of the most important cases is the three-dimensional O(N) UC \cite{PV-02}.
In the presence of a 2D surface, the scenario above is realized for $N=1$ (Ising) and $N=2$ ($XY$) cases. Surface critical behavior for the Heisenberg UC is instead not yet fully understood.
Experiments have proven the realization of the ordinary surface UC for Gd samples at its bulk critical point, in the O(3) UC \cite{AP-00}.
Since the Mermin-Wagner-Hohenberg theorem \cite{MW-66,*MW-66_erratum,Hohenberg-67,Halperin-19} forbids a surface transition, one could conclude that only the ordinary UC is realized. While early Monte Carlo (MC) simulations supported this picture \cite{Krech-00}, a later MC study claimed a possible Berezinskii-Kosterlitz-Thouless- (BKT) like surface transition \cite{DBN-05}.
This problem has recently attracted renewed attention in the context of quantum critical behavior, where several investigations reported puzzling results.
MC simulations of dimerized spin-$1/2$ systems, exhibiting a classical Heisenberg bulk UC, have found nonordinary surface exponents for some geometrical settings \cite{SS-12,ZW-17,DZG-18,WPTW-18}. Such a novel behavior has been attributed to a relevant topological $\theta$ term at the boundary, which is irrelevant for the bulk critical behavior \cite{WPTW-18}. A theory for a direct transition between a N\'eel and a valence-bond solid (VBS) in nonlocal 1D quantum systems has been put forward to explain the observed behavior \cite{JXWX-20}.
Nevertheless, quite remarkably a MC study of a dimerized $S=1$ system reported a surface critical exponent close (although not identical) to that of the $S=1/2$ case \cite{WW-19}, whereas VBS correlations decay faster than for the $S=1/2$ case \cite{WW-20}. Similar exponents have been found at the boundary of coupled Haldane chains \cite{ZDZG-20}.
For a $S=1$ system a topological $\theta$ term is absent, and so via a standard quantum-to-classical mapping \cite{Sachdev-book} it should correspond to a classical 3D O(3) model with a surface.
It is therefore unclear whether a boundary $\theta$ term is responsible for the observed nonordinary exponents for $S=1/2$ systems.
In this context, a recent field-theoretical study has put forward different possible scenarios for the surface transition in the Heisenberg UC \cite{Metlitski-20}, the realization of which depends on the values of some amplitudes at the so-called normal surface UC \cite{Diehl-94,BD-94,Binder-83,Diehl-86,Pleimling-review}.
Motivated by these developments, and by the need to understand the classical surface O(3) UC in 3D, we investigate here an improved lattice model by means of MC simulations. By tuning a surface coupling we unveil the existence of a boundary phase transition, separating the ordinary and extraordinary phases.
Our findings provide an explanation for abovementioned results.

{\it Model.}---We simulate the $\phi^4$ model, defined on a 3D $L_\parallel\times L_\parallel\times L$ lattice, with periodic boundary conditions (BCs) on directions corresponding to $L_\parallel$, and open BCs on the remaining direction. The reduced Hamiltonian, such that the Gibbs weight is $\exp(-\cal H)$, is
\begin{equation}
  \begin{split}
    {\cal H} = &-\beta\sum_{\< i\ j\>}\vec{\phi}_i\cdot\vec{\phi}_j
    -\beta_{s,\downarrow}\sum_{\< i\ j\>_{s\downarrow}}\vec{\phi}_i\cdot\vec{\phi}_j\\
    &-\beta_{s,\uparrow}\sum_{\< i\ j\>_{s\uparrow}}\vec{\phi}_i\cdot\vec{\phi}_j
    +\sum_i[\vec{\phi}_i^{\,2}+\lambda(\vec{\phi}_i^{\,2}-1)^2],
  \end{split}
  \label{model}
\end{equation}
where $\vec{\phi}_x$ is a three-components real field on the lattice site $x$, the first sum extends over the nearest-neighbor pairs where at least one field belongs to the inner bulk, the second and third sums pertain to the lower and upper surface, and the last term is summed over all lattice sites. 

For $\lambda\rightarrow\infty$, the Hamiltonian (\ref{model}) reduces to the classical O(3) model. In the $(\beta, \lambda)$ plane, the bulk exhibits a second-order transition line in the Heisenberg UC \cite{CHPRV-02,PV-02}. At $\lambda=5.17(11)$ the model is {\it improved} \cite{Hasenbusch-20}, i.e., leading bulk scaling corrections $\propto L^{-\omega_1}$, $\omega_1=0.759(2)$, are suppressed and those due to the next-to-leading irrelevant bulk operator decay fast as $L^{-\omega_2}$, $\omega_2\approx 2$ \cite{NR-84}. Additional corrections to scaling originate from the presence of surfaces.
Improved lattice models are instrumental in high-precision MC simulations \cite{PV-02}, and in particular in boundary critical phenomena
\cite{Hasenbusch-09b,Hasenbusch-10c,PTD-10,Hasenbusch-11,Hasenbusch-11b,Hasenbusch-12,PTTD-13,PT-13,PTTD-14,PTAW-17}.
For $\lambda=5.2$, the model is critical at $\beta=\numprint{0.68798521}(8)$ \cite{Hasenbusch-20}. The couplings $\beta_{s,\downarrow}$, $\beta_{s,\uparrow}$ control the surface enhancement of the order parameter.
Here we fix $L_\parallel=L$, $\lambda=5.2$, $\beta=\numprint{0.68798521}$, $\beta_{s,\downarrow}=\beta_{s,\uparrow}=\beta_s$ and study the surface critical behavior on varying $\beta_s$. We compute improved estimators of surface observables by averaging them over the two surfaces.
MC simulations are performed by combining Metropolis, overrelaxation, and Wolff single-cluster updates \cite{Wolff-89,SM}.

{\it Special transition.}---For $\beta_s=\beta$ there is no surface enhancement and at the bulk critical point the model realizes the ordinary UC. Its critical behavior will be studied elsewhere \cite{PT-23}.
To investigate the surface critical behavior we proceed in two steps. We first analyze RG-invariant quantities, with the aim of locating the onset of a phase transition, and determine the fixed-point values.
Then, we employ these results in a finite-size scaling (FSS) \cite{Privman-90} analysis to compute universal critical exponents.
In the vicinity of a surface transition at $\beta_s=\beta_{s,c}$, and neglecting for the moment scaling corrections, a RG-invariant observable $R$ satisfies
\begin{equation}
  R = f((\beta_s-\beta_{s,c})L^{y_{\rm sp}}),
  \label{FSS_RGinv}
\end{equation}
where $y_{\rm sp}$ is the scaling dimension of the relevant scaling field associated with the transition.
We consider the surface Binder ratio $U_4$:
\begin{equation}
  U_4 \equiv \frac{\<(\vec{M}_s^2)^2\>}{\<\vec{M}_s^2\>^2}, \qquad \vec{M}_s\equiv \sum_{i\in\text{surface}}\vec{\phi}_i.
  \label{U4}
\end{equation}
\begin{figure}[t]
  \centering
  \includegraphics[width=0.85\linewidth]{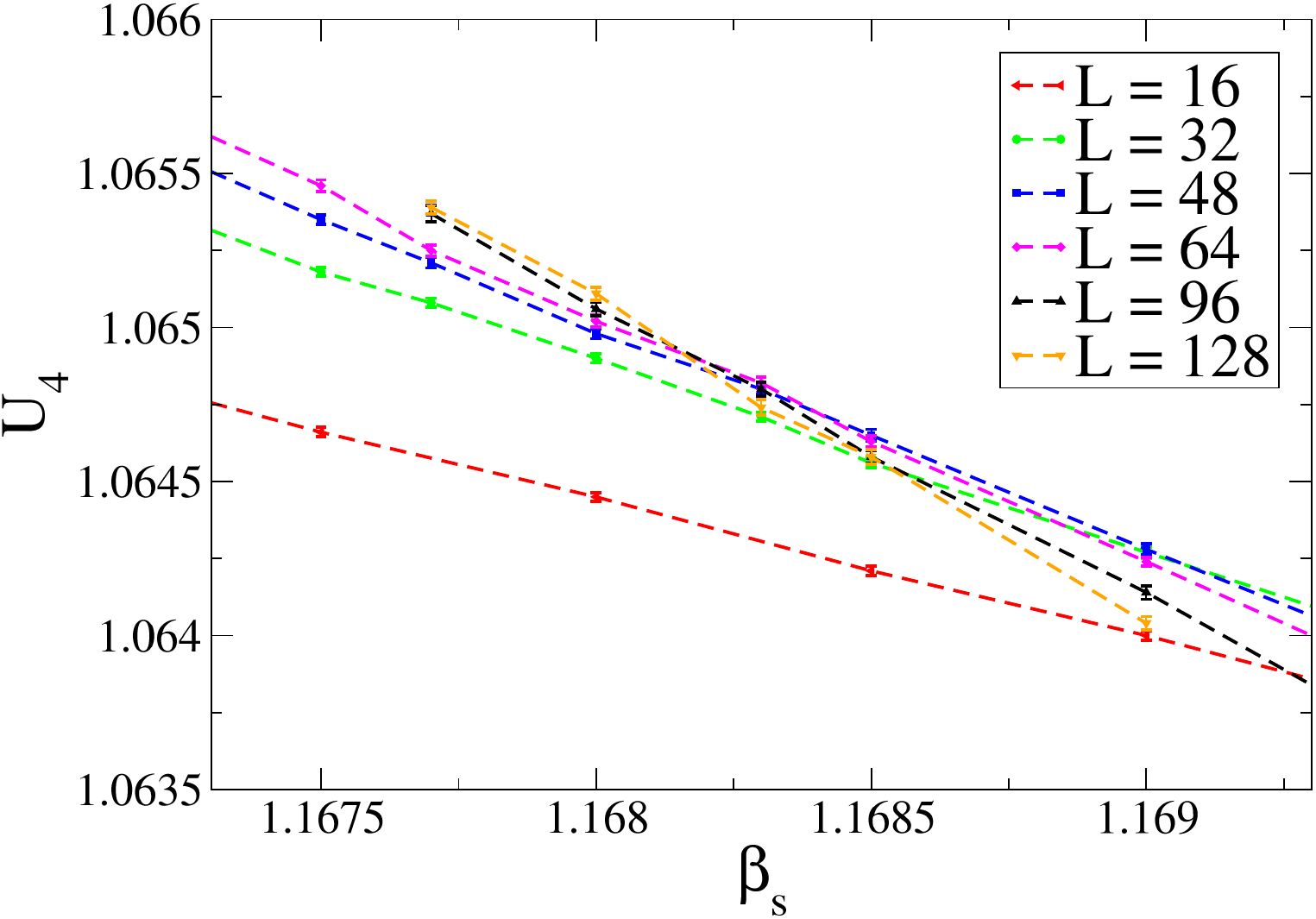}
  \caption{Plot of the RG-invariant quantity $U_4$ defined in Eq.~(\ref{U4}) as a function of $\beta_s$. MC error bars \cite{CS-86,MS-88,Sokal_lecture,Young_notes} are $\approx 10^{-5}$.}
  \label{fig.U4}
\end{figure}

In Fig.~\ref{fig.U4} we show $U_4$ as function of $\beta_s$ for lattice sizes $L=16$, $32$, $48$, $64$, $96$, $128$.
We observe a crossing indicating a surface phase transition.
Its existence is visually more evident when data are plotted on a larger scale \cite{SM}.
The slope of $U_4$ appears to increase rather slowly with $L$, such that a rather high precision in the MC data ($\approx 10^{-5}$) is needed in order to show the crossing. 
Within such a high accuracy, scaling corrections are visible, although for instance the data for $L=16$ deviate by a mere $\lesssim 0.1\%$ from the data at $L=64$.
For a quantitative determination of critical parameters, we expand the right-hand side of Eq.~(\ref{FSS_RGinv}) in Taylor series \cite{PTPV-09}, including possible scaling corrections, as:
\begin{equation}
  \begin{split}
    R = R^* &+ \sum_{n=1}^m a_n(\beta_s-\beta_{s,c})^nL^{ny_{\rm sp}} \\
    &+ L^{-\omega}\sum_{n=0}^k b_n(\beta_s-\beta_{s,c})^nL^{ny_{\rm sp}},
  \end{split}
  \label{FSS_RGinv_taylor}
\end{equation}
where $\omega$ is the leading correction-to-scaling exponent. We first consider fits of $R=U_4$ neglecting scaling corrections and for $m=1$.
\begin{table}[b]
  \caption{Fits of $R=U_4$ to the right-hand side of Eq.~(\ref{FSS_RGinv_taylor}), with $m=1$, neglecting scaling corrections $\propto L^{-\omega}$ (above), and including corrections to scaling with $\omega=1$ and $k=0$ (below).
  }
  \begin{ruledtabular}
    \begin{tabular}{l@{\hspace{2em}}.@{\hspace{2em}}.@{\hspace{2em}}.q}
      \multicolumn{1}{l}{$L_{\rm min}$} & \multicolumn{1}{c}{$U_4^*$} & \multicolumn{1}{c}{$\beta_{s,c}$} & \multicolumn{1}{c}{$y_{\rm sp}$} & \multicolumn{1}{l}{$\chi^2/{\rm d.o.f.}$} \\
      \hline
      $16$ &  1.06385(5) & 1.16941(6) & 0.27(2)  & 50.2 \\
      $32$ &  1.06463(2) & 1.16847(3) & 0.40(2)  & 3.9 \\
      $48$ &  1.06481(3) & 1.16827(3) & 0.40(3)  & 1.0 \\
      $64$ &  1.06487(4) & 1.16821(5) & 0.39(4)  & 1.0 \\
      $96$ &  1.0649(2)  & 1.1681(2)  & 0.36(11) & 0.9 \\
      \hline
      $16$ &  1.06557(5) & 1.16764(5) & 0.40(2) & 1.0 \\
      $32$ &  1.0654(1)  & 1.16779(9) & 0.39(2) & 0.8\\
    \end{tabular}
  \end{ruledtabular}
  \label{fits_U4}
\end{table}
Corresponding results are reported in Table \ref{fits_U4}, as a function of the minimum lattice size $L_{\rm min}$ taken into account. Results are overall stable, exhibiting however a small detectable drift on increasing $L_{\rm min}$, which is larger than the statistical accuracy of the fit. Furthermore, a good $\chi^2/{\rm d.o.f.}$ (${\rm d.o.f.}$ denotes the degrees of freedom) is found only for $L_{\rm min} \ge 48$.
In line with the above observation on the slope of $U_4$, the fitted value of $y_{\rm sp}$ is unusually small.
Increasing $m$ to $2$ does not change significantly $\chi^2/{\rm d.o.f.}$, indicating that the approximation $m=1$ is adequate \cite{SM}.
The small value of $y_{\rm sp}$ can potentially result in slowly decaying analytical scaling corrections $\propto L^{-y_{\rm sp}}$, originating from nonlinearities in the scaling field \cite{AF-83}. To check their relevance, we have repeated the fits including a quadratic correction to the relevant scaling field $(\beta_s-\beta_{s,c})\rightarrow (\beta_s-\beta_{s,c}) + B(\beta_s-\beta_{s,c})^2$. We obtain identical results, and the fitted values of $B$ vanish within error bars, therefore analytical scaling corrections are negligible for the range of data in exam \cite{SM}.
Fits including the term $\propto L^{-\omega}$, with a free $\omega$ parameter, are consistent with $\omega\gtrsim 1$ \cite{SM}. Since a correction term $\propto L^{-1}$ is in any case expected for nonperiodic BCs \cite{CF-76,CPV-14}, we can safely assume that leading scaling corrections are $\propto L^{-1}$.
To obtain more accurate results,
we have repeated the fits to Eq.~(\ref{FSS_RGinv_taylor}) setting $\omega=1$ and $k=0$. Corresponding results reported in Table \ref{fits_U4} are stable, with a good $\chi^2/{\rm d.o.f.}$ By judging conservatively the variation of estimates, we obtain the critical-point value of $U_4^* = 1.0652(4)$.
We use this result to evaluate critical exponents with the method of FSS at fixed phenomenological coupling \cite{Hasenbusch-99,PT-11}. This technique consists in an analysis of MC data done by fixing the value of a RG-invariant observable $R$ (here, $R=U_4$), thereby trading the fluctuations of $R$ with fluctuations of a parameter driving the transition (here, $\beta_s$). This method has been used in several high-precision MC studies of critical phenomena \cite{HPTPV-07,Hasenbusch-10,Hasenbusch-19,Hasenbusch-20}, and can lead to significant gains in the error bars \cite{HPTPV-07,PT-11}. A discussion of the method can be found in Ref.~\cite{PT-11}.
For this analysis we have complemented MC data shown in Fig.~\ref{fig.U4} with an additional simulation at $L=192$.
To compute the exponent $y_{\rm sp}$, we consider derivatives of a RG-invariant observable $R$ with respect to $\beta_s$, at fixed $U_4= 1.0652$. According to FSS, and including leading $L^{-1}$ scaling corrections,
\begin{equation}
  \frac{dR}{d\beta_s} = AL^{y_{\rm sp}} \left(1+BL^{-1}\right).
\label{FSS_dR_dbetas}
\end{equation}
We consider $R=U_4$ and the ratio $R=Z_a/Z_p$ of the partition function with antiperiodic and periodic BCs on a direction parallel to the surfaces, sampled with the boundary-flip algorithm \cite{Hasenbusch-93,CHPRV-01}.
\begin{table}
  \caption{Fits of $dR/d\beta_s$ to Eq.~(\ref{FSS_dR_dbetas}) for $R=U_4$ and $R=Z_a/Z_p$ at fixed $U_4^*=1.0652$.
    Fits above are obtained setting $B=0$ in Eq.~(\ref{FSS_dR_dbetas}), i.e., neglecting scaling corrections, fits below include the term $BL^{-1}$.}
  \begin{ruledtabular}
    \begin{tabular}{ll@{}.@{}Q}
      \multicolumn{1}{l}{Observable} & \multicolumn{1}{c}{$L_{\rm min}$} & \multicolumn{1}{c}{$y_{\rm sp}$} & \multicolumn{1}{c}{$\chi^2/{\rm d.o.f.}$} \\
      \hline
                & $16$ &  0.3952(7) & 37.9 \\
                & $32$ &  0.381(2)  & 4.7 \\
$dU_4/d\beta_s$ & $48$ &  0.374(2)  & 0.2 \\
                & $64$ &  0.372(4)  & 0.2 \\
                & $96$ &  0.369(6)  & 0.03 \\[1em]
                      & $16$ &  0.364(3) & 0.8 \\
                      & $32$ &  0.362(5) & 1.0 \\
$d(Z_a/Z_p)/d\beta_s$ & $48$ &  0.364(9) & 1.3 \\
                      & $64$ &  0.35(2)  & 0.01 \\
                      & $96$ &  0.34(3)  & 0.03 \\
      \hline
                & $16$ & 0.361(3)  & 0.4 \\
$dU_4/d\beta_s$ & $32$ & 0.357(6)  & 0.3 \\
                & $48$ & 0.366(11) & 0.07 \\[1em]
                      & $16$ &  0.36(1) &  1.0 \\
$d(Z_a/Z_p)/d\beta_s$ & $32$ &  0.35(2) &  1.3 \\
                     & $48$ &  0.29(4) &  0.4 \\
    \end{tabular}
  \end{ruledtabular}
  \label{fits_ysp}
\end{table}
In Table \ref{fits_ysp} we report the  various results of fits to Eq.~(\ref{FSS_dR_dbetas}). By looking conservatively at the variation of the results, we estimate
\begin{equation}
  y_{\rm sp} = 0.36(1), \qquad \nu_{\rm sp}\equiv 1/y_{\rm sp} = 2.78(8).
  \label{ysp}
\end{equation}
This result also agrees with the less precise fits shown in Table \ref{fits_U4}.
To compute the surface magnetic exponent $\eta_\parallel$ we measure the surface susceptibility
\begin{equation}
  \chi_s = \frac{1}{L^2}\sum_{i,j\in {\rm surface}} \vec{\phi}_i\cdot\vec{\phi}_j.
  \label{chidef}
\end{equation}
In agreement with standard surface FSS \cite{Binder-83}, we fit MC data for $\chi_s$ at fixed $U_4^*$ to
\begin{equation}
  \chi_s = AL^{1-\eta_\parallel}\left(1+BL^{-1}\right),
  \label{FSS_chis}
\end{equation}
where as above we allow for a correction-to-scaling term $\propto L^{-1}$. Fit results are reported in Table \ref{fits_eta}. We estimate
\begin{equation}
  \eta_\parallel = -0.473(2).
  \label{eta}
\end{equation}

\begin{table}[b]
  \caption{Fits of $\chi_s$ at fixed $U_4=1.0652$ to the right-hand side of Eq.~(\ref{FSS_chis}) neglecting the scaling corrections $\propto L^{-1}$ (above), and including them (below).}
  \begin{ruledtabular}
    \begin{tabular}{l@{}.@{}q}
      \multicolumn{1}{c}{$L_{\rm min}$} & \multicolumn{1}{c}{$\eta_\parallel$} & \multicolumn{1}{c}{$\chi^2/{\rm d.o.f.}$} \\
      \hline
      $16$ &   -0.47760(7) & 146.9 \\
      $32$ &   -0.4753(1)  & 12.3 \\
      $48$ &   -0.4746(2)  & 3.1 \\
      $64$ &   -0.4742(2)  & 1.4 \\
      $96$ &   -0.4736(4)  & 0.2 \\
      \hline
      $16$ &   -0.4721(2)  & 0.4 \\
      $32$ &   -0.4725(4)  & 0.2 \\
      $48$ &   -0.4723(8)  & 0.3 \\
    \end{tabular}
  \end{ruledtabular}
  \label{fits_eta}
\end{table}
We checked that
varying the fixed value $U_4^* = 1.0652(4)$ within one error bar gives negligible variations in the resulting critical exponents \cite{SM}.
Finally, FSS at fixed $U_4^*$ allows us to estimate $\beta_{s,c} = 1.1678(2)$ \cite{SM}.

\begin{figure}
  \centering
  \includegraphics[width=\linewidth]{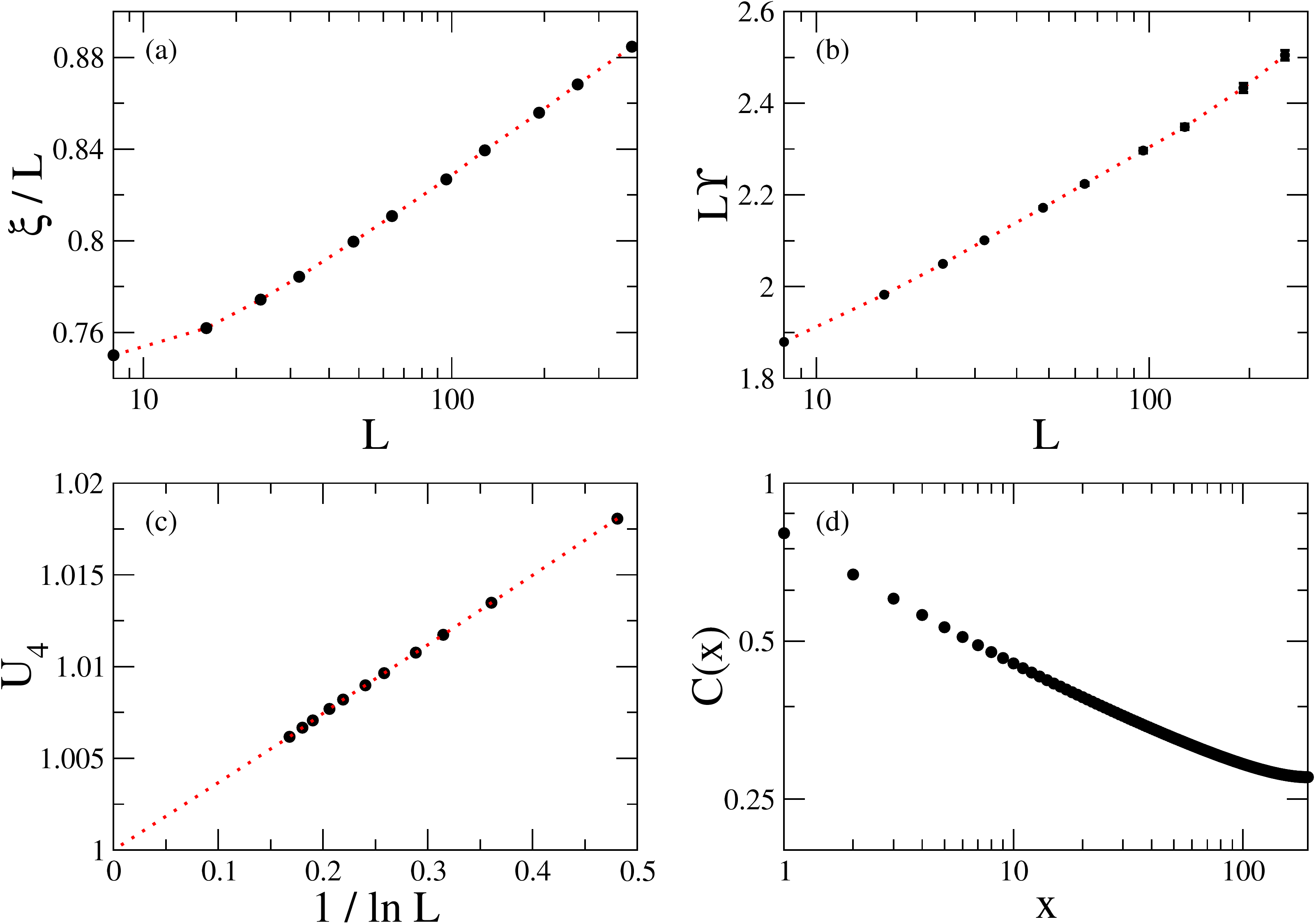}
  \caption{Observables for $\beta_s=1.5$, in the extraordinary phase. The ratio $\xi/L$ (a) and $\Upsilon L$ (b) in semilogarithmic scale. (c) The surface Binder ratio $U_4$ as a function of $1/\ln L$.
    Dotted lines are a guide to the eye. (d) The surface correlations of the order parameter for $L=348$.
    When not visible, statistical error bars are of the order or smaller than the point size.
  }
  \label{fig.extra}
\end{figure}
{\it Extraordinary phase.}---The existence of a surface phase transition implies an extraordinary phase for $\beta_s>\beta_{s,c}$. To investigate it, we have simulated the model at $\beta_s=1.5$, for lattice sizes $8\le L \le 384$.
In Figs.~\ref{fig.extra}(a) and \ref{fig.extra}(b) we plot the ratio $\xi/L$ of the surface correlation length $\xi$ \footnote{See Appendix A of Ref.~\cite{PTHAH-14} for a discussion on the definition of $\xi$ for a finite lattice, as well as the SM \cite{SM}.} over the lattice size $L$, and the product $\Upsilon L$, where $\Upsilon$ is the helicity modulus \cite{FBJ-73,Metlitski-helicity}. Both quantities exhibit a logarithmic growth with $L$, indicating a violation of standard FSS. The surface Binder ratio $U_4$ shown in Fig.~\ref{fig.extra}(c) is rather close to $1$, and exhibits a logarithmic approach to $1$.
Nevertheless, the surface is not ordered: its two-point function $C(x){\equiv} \<\vec{\phi}_0\cdot\vec{\phi}_x\>$ for the largest lattice size $L=384$ shown in Fig.~\ref{fig.extra}(d) exhibits a slow, visible decay.
Furthermore, for an ordered surface, $\xi/L \ {\sim}\  L$ and $\Upsilon \ {\sim}\  \text{const}$, in contrast with Figs.~\ref{fig.extra}(a) and \ref{fig.extra}(b).
These findings support the scenario of a so-called ``extra-ordinary-log'' phase, recently put forward in Ref.~\cite{Metlitski-20}. In such a phase,
$C(x{\rightarrow}\infty)\ {\propto}\ \ln(x)^{-q}$, where $q$ is a universal exponent determined by some amplitudes in the normal UC.
Fits of $C(L/2)$, $C(L/4)$ to $\ln(L/l_0)^{-q}$, and of $\chi$ to $L^2 \ln(L/l_0)^{-q}$ \cite{Metlitski-private}, provide an estimate of $q\simeq 2.1(2)$ \cite{SM}.
Moreover, in the ``extra-ordinary-log'' phase $U_4-1\ {\propto}\ (\ln L)^{-2}$, $(\xi/L)^2\simeq (\alpha/2)\ln(L)$ and $\Upsilon L  \simeq (2/3)\alpha\ln(L)$, for $L {\rightarrow} \infty$, with $\alpha=1/(\pi q)$ a universal RG parameter \cite{Metlitski-private,PTKM-24}.
Indeed, fits of $(\xi/L)^2$ to $(\alpha/2)\ln L + B$ give $\alpha\approx 0.14$, showing however some drift in the estimate as a function of the minimum lattice size taken into account.
Such a value is nevertheless consistent with
the estimate of $q$ reported above, which corresponds to $\alpha\simeq 0.15(2)$.
Corresponding fits of $L\Upsilon$ give less stable results.
Judging from the trends in the fit results, one can conclude $\alpha \gtrsim 0.16$, again consistent with previous estimates \cite{erratum}.
We stress that error bars reported above should be taken with some grain of salt, since they stem from fits that neglect subleading corrections; these are likely to be important, as illustrated, e.g., by other critical models with marginal perturbations \cite{HPTPV-08b}.
A more quantitative precise assessment of the extraordinary phase is outside the scope of the present work.

{\it Discussion.}---In this work we have elucidated the boundary critical behavior of the classical 3D O(3) UC, in the presence of a 2D surface.
A previous MC study, assuming the existence of the ordinary UC only, did not consider RG-invariant observables and reported just a crossover to the ordinary UC for a strong enough surface enhancement \cite{Krech-00}. A later study observed a flattening in the curves of the RG-invariant $Q_{11}\equiv 1/U_4$ for large enough surface coupling, and interpreted this as the onset of a BKT-like transition, without further investigations \cite{DBN-05}.
Here, by means of large-statistics MC simulations of an improved model, where leading scaling corrections are suppressed, and a quantitative FSS analysis, we have proven the existence of a standard special phase transition, with an unusually small, but finite, leading relevant exponent. The extraordinary phase displays 
slowly decaying correlations and, remarkably, a logarithmic violation of FSS, supportive of the ``extra-ordinary-log'' scenario of Ref.~\cite{Metlitski-20}.
A comprehensive theory of such a rather uncommon FSS violation is presently unavailable; hopefully, this work will stimulate research in this direction.
 These findings also provide an explanation to recent MC results on the boundary critical behavior of quantum spin models \cite{SS-12,ZW-17,DZG-18,WPTW-18,WW-19,WW-20,ZDZG-20}. The exponent $\eta_\parallel$ found for some geometrical settings is close to that of the special transition, Eq.~(\ref{eta}), thus suggesting that those quantum spin models are ``accidentally'' close to the special transition.
The observed $\eta_\parallel$ is also close to a simple evaluation of the two-loops $\varepsilon$-expansion series \cite{DD-81c,Diehl-86,CGLT-83,KNSFCL-91,*KNSFCL-91_erratum} by setting $\varepsilon\ {=}\ 1$ and $N\ {=}\ 3$ \cite{DZG-18}. However, the $\varepsilon$-expansion result for $y_{\rm sp}$ differs significantly from Eq.~(\ref{ysp}) \cite{SM}.
Generally, the realization of the special UC requires a fine-tuning of boundary couplings, because the corresponding fixed point is unstable.
Nevertheless, the unusually small value of $y_{\rm sp}$ [Eq.~(\ref{ysp})] implies a slow crossover from the special fixed point when the model is tuned away from the special transition.
In other words, a small $y_{\rm sp}$ results in
a (relatively) large region, $(\beta_s-\beta_{s,c})L^{y_{\rm sp}}=O(1)$, where FSS is controlled by the special fixed point and the observed exponents are close to those of the special UC, without the need of a fine-tuning.
This plausibly explains at least the results for $S=1$ quantum models of Refs.~\cite{WW-19,ZDZG-20}, where a topological $\theta$ term is absent. Also, we observe that the exponent $\eta_\parallel$ reported in Refs.~\cite{WW-19,ZDZG-20} deviates for about $15\%$ from $\eta_\parallel$ at the special point [Eq.~(\ref{eta})], suggesting that the models are not exactly at the special transition.
Concerning the $S=1/2$ case, we notice that the small value of $y_{\rm sp}$ implies that the special fixed point is located at a small, possibly perturbatively accessible, value of the coupling constant $g^*$ of the field theory studied in Ref.~\cite{Metlitski-20}.
Accordingly, if the special transition occurs in the presence of VBS order, $\eta_\parallel$ is expected to be identical to the $S=1$ case, whereas for a direct magnetic-VBS transition, as advocated in Ref.~\cite{JXWX-20}, nonperturbative corrections to $\eta_\parallel$ due to the topological $\theta$ term are expected to be small \cite{Metlitski-20}. This would explain the similarity of the $\eta_\parallel$ exponent in dimerized $S=1/2$ models \cite{ZW-17,DZG-18,WPTW-18} with that of the special transition [Eq.~(\ref{eta})].
Finally, to close the loop, it would be highly desirable to investigate the boundary critical behavior of quantum spin models with a tunable surface coupling, such as those considered in Refs.~\cite{WPTW-18,WW-19}, so as to detect a surface phase transition and compare with the present findings.

The author is grateful to Max Metlitski for insightful discussions and useful communications on the manuscript. The author thanks Stefan Wessel for useful comments on the manuscript. F.P.T. is funded by the Deutsche Forschungsgemeinschaft (DFG, German Research Foundation)--Project No. 414456783.
The author gratefully acknowledges the Gauss Centre for Supercomputing e.V. for funding this project by providing computing time through the John von Neumann Institute for Computing (NIC) on the GCS Supercomputer JUWELS at Jülich Supercomputing Centre (JSC) \cite{JUWELS}.

\bibliography{francesco,extra}

\begin{thebibliography}{76}%
\makeatletter
\providecommand \@ifxundefined [1]{%
 \@ifx{#1\undefined}
}%
\providecommand \@ifnum [1]{%
 \ifnum #1\expandafter \@firstoftwo
 \else \expandafter \@secondoftwo
 \fi
}%
\providecommand \@ifx [1]{%
 \ifx #1\expandafter \@firstoftwo
 \else \expandafter \@secondoftwo
 \fi
}%
\providecommand \natexlab [1]{#1}%
\providecommand \enquote  [1]{``#1''}%
\providecommand \bibnamefont  [1]{#1}%
\providecommand \bibfnamefont [1]{#1}%
\providecommand \citenamefont [1]{#1}%
\providecommand \href@noop [0]{\@secondoftwo}%
\providecommand \href [0]{\begingroup \@sanitize@url \@href}%
\providecommand \@href[1]{\@@startlink{#1}\@@href}%
\providecommand \@@href[1]{\endgroup#1\@@endlink}%
\providecommand \@sanitize@url [0]{\catcode `\\12\catcode `\$12\catcode
  `\&12\catcode `\#12\catcode `\^12\catcode `\_12\catcode `\%12\relax}%
\providecommand \@@startlink[1]{}%
\providecommand \@@endlink[0]{}%
\providecommand \url  [0]{\begingroup\@sanitize@url \@url }%
\providecommand \@url [1]{\endgroup\@href {#1}{\urlprefix }}%
\providecommand \urlprefix  [0]{URL }%
\providecommand \Eprint [0]{\href }%
\providecommand \doibase [0]{https://doi.org/}%
\providecommand \selectlanguage [0]{\@gobble}%
\providecommand \bibinfo  [0]{\@secondoftwo}%
\providecommand \bibfield  [0]{\@secondoftwo}%
\providecommand \translation [1]{[#1]}%
\providecommand \BibitemOpen [0]{}%
\providecommand \bibitemStop [0]{}%
\providecommand \bibitemNoStop [0]{.\EOS\space}%
\providecommand \EOS [0]{\spacefactor3000\relax}%
\providecommand \BibitemShut  [1]{\csname bibitem#1\endcsname}%
\let\auto@bib@innerbib\@empty
\bibitem [{\citenamefont {Dosch}(2006)}]{Dosch-book}%
  \BibitemOpen
  \bibfield  {author} {\bibinfo {author} {\bibfnamefont {H.}~\bibnamefont
  {Dosch}},\ }\href {https://books.google.de/books?id=yZl0DgAAQBAJ} {\emph
  {\bibinfo {title} {Critical Phenomena at Surfaces and Interfaces: Evanescent
  X-Ray and Neutron Scattering}}},\ Springer Tracts in Modern Physics\
  (\bibinfo  {publisher} {Springer Berlin Heidelberg},\ \bibinfo {year}
  {2006})\BibitemShut {NoStop}%
\bibitem [{\citenamefont {{Binder}}(1983)}]{Binder-83}%
  \BibitemOpen
  \bibfield  {author} {\bibinfo {author} {\bibfnamefont {K.}~\bibnamefont
  {{Binder}}},\ }\bibfield  {title} {\bibinfo {title} {Critical behavior at
  surfaces},\ }in\ \href@noop {} {\emph {\bibinfo {booktitle} {Phase
  Transitions and Critical Phenomena}}},\ \bibinfo {series} {Phase Transitions
  and Critical Phenomena}, Vol.~\bibinfo {volume} {8},\ \bibinfo {editor}
  {edited by\ \bibinfo {editor} {\bibfnamefont {C.}~\bibnamefont {{Domb}}}\
  and\ \bibinfo {editor} {\bibfnamefont {J.~L.}\ \bibnamefont {{Lebowitz}}}}\
  (\bibinfo  {publisher} {Academic Press},\ \bibinfo {address} {London},\
  \bibinfo {year} {1983})\ p.~\bibinfo {pages} {1}\BibitemShut {NoStop}%
\bibitem [{\citenamefont {{Diehl}}(1986)}]{Diehl-86}%
  \BibitemOpen
  \bibfield  {author} {\bibinfo {author} {\bibfnamefont {H.~W.}\ \bibnamefont
  {{Diehl}}},\ }\bibfield  {title} {\bibinfo {title} {Field-theoretical
  approach to critical behaviour at surfaces},\ }in\ \href@noop {} {\emph
  {\bibinfo {booktitle} {Phase Transitions and Critical Phenomena}}},\ \bibinfo
  {series} {Phase Transitions and Critical Phenomena}, Vol.~\bibinfo {volume}
  {10},\ \bibinfo {editor} {edited by\ \bibinfo {editor} {\bibfnamefont
  {C.}~\bibnamefont {{Domb}}}\ and\ \bibinfo {editor} {\bibfnamefont {J.~L.}\
  \bibnamefont {{Lebowitz}}}}\ (\bibinfo  {publisher} {Academic Press},\
  \bibinfo {address} {London},\ \bibinfo {year} {1986})\ p.~\bibinfo {pages}
  {75}\BibitemShut {NoStop}%
\bibitem [{\citenamefont {{Pleimling}}(2004)}]{Pleimling-review}%
  \BibitemOpen
  \bibfield  {author} {\bibinfo {author} {\bibfnamefont {M.}~\bibnamefont
  {{Pleimling}}},\ }\bibfield  {title} {\bibinfo {title} {{Critical phenomena
  at perfect and non-perfect surfaces}},\ }\href
  {https://doi.org/10.1088/0305-4470/37/19/R01} {\bibfield  {journal} {\bibinfo
   {journal} {\JPAOLD}\ }\textbf {\bibinfo {volume} {37}},\ \bibinfo {pages}
  {R79} (\bibinfo {year} {2004})},\ \Eprint
  {https://arxiv.org/abs/cond-mat/0402574} {cond-mat/0402574} \BibitemShut
  {NoStop}%
\bibitem [{\citenamefont {{Cardy}}(1996)}]{Cardy-book}%
  \BibitemOpen
  \bibfield  {author} {\bibinfo {author} {\bibfnamefont {J.}~\bibnamefont
  {{Cardy}}},\ }\href@noop {} {\emph {\bibinfo {title} {{Scaling and
  Renormalization in Statistical Physics}}}}\ (\bibinfo  {publisher} {Cambridge
  University Press},\ \bibinfo {address} {Cambridge},\ \bibinfo {year}
  {1996})\BibitemShut {NoStop}%
\bibitem [{\citenamefont {Fisher}\ and\ \citenamefont
  {de~Gennes}(1978)}]{FG-78}%
  \BibitemOpen
  \bibfield  {author} {\bibinfo {author} {\bibfnamefont {M.~E.}\ \bibnamefont
  {Fisher}}\ and\ \bibinfo {author} {\bibfnamefont {P.-G.}\ \bibnamefont
  {de~Gennes}},\ }\bibfield  {title} {\bibinfo {title} {{Ph\'enom\`enes aux
  parois dans un m\'elange binaire critique}},\ }\href
  {http://gallica.bnf.fr/ark:/12148/bpt6k62353730/f61.image.r=fisher%20de%20gennes.langEN}
  {\bibfield  {journal} {\bibinfo  {journal} {C.~R.~Acad.~Sci.~Paris Ser.~B~}\
  }\textbf {\bibinfo {volume} {287}},\ \bibinfo {pages} {207} (\bibinfo {year}
  {1978})}\BibitemShut {NoStop}%
\bibitem [{\citenamefont {Krech}(1994)}]{Krech-94}%
  \BibitemOpen
  \bibfield  {author} {\bibinfo {author} {\bibfnamefont {M.}~\bibnamefont
  {Krech}},\ }\href {http://books.google.de/books?id=0ZuCngEACAAJ} {\emph
  {\bibinfo {title} {The Casimir Effect in Critical Systems}}}\ (\bibinfo
  {publisher} {World Scientific},\ \bibinfo {address} {London},\ \bibinfo
  {year} {1994})\BibitemShut {NoStop}%
\bibitem [{\citenamefont {{Krech}}(1999)}]{Krech-99}%
  \BibitemOpen
  \bibfield  {author} {\bibinfo {author} {\bibfnamefont {M.}~\bibnamefont
  {{Krech}}},\ }\bibfield  {title} {\bibinfo {title} {{Fluctuation-induced
  forces in critical fluids}},\ }\href
  {https://doi.org/10.1088/0953-8984/11/37/201} {\bibfield  {journal} {\bibinfo
   {journal} {\JPCM}\ }\textbf {\bibinfo {volume} {11}},\ \bibinfo {pages}
  {R391} (\bibinfo {year} {1999})},\ \Eprint
  {https://arxiv.org/abs/cond-mat/9909413} {cond-mat/9909413} \BibitemShut
  {NoStop}%
\bibitem [{\citenamefont {{Gambassi}}(2009)}]{Gambassi-09}%
  \BibitemOpen
  \bibfield  {author} {\bibinfo {author} {\bibfnamefont {A.}~\bibnamefont
  {{Gambassi}}},\ }\bibfield  {title} {\bibinfo {title} {{The Casimir effect:
  From quantum to critical fluctuations}},\ }\href
  {https://doi.org/10.1088/1742-6596/161/1/012037} {\bibfield  {journal}
  {\bibinfo  {journal} {J. Phys.: Conf. Ser.}\ }\textbf {\bibinfo {volume}
  {161}},\ \bibinfo {eid} {012037} (\bibinfo {year} {2009})},\ \Eprint
  {https://arxiv.org/abs/0812.0935} {arXiv:0812.0935 [cond-mat.stat-mech]}
  \BibitemShut {NoStop}%
\bibitem [{\citenamefont {{Gambassi}}\ and\ \citenamefont
  {{Dietrich}}(2011)}]{GD-11}%
  \BibitemOpen
  \bibfield  {author} {\bibinfo {author} {\bibfnamefont {A.}~\bibnamefont
  {{Gambassi}}}\ and\ \bibinfo {author} {\bibfnamefont {S.}~\bibnamefont
  {{Dietrich}}},\ }\bibfield  {title} {\bibinfo {title} {{Critical Casimir
  forces steered by patterned substrates}},\ }\href
  {https://doi.org/10.1039/c0sm00635a} {\bibfield  {journal} {\bibinfo
  {journal} {Soft Matter}\ }\textbf {\bibinfo {volume} {7}},\ \bibinfo {pages}
  {1247} (\bibinfo {year} {2011})},\ \Eprint {https://arxiv.org/abs/1011.1831}
  {arXiv:1011.1831 [cond-mat.soft]} \BibitemShut {NoStop}%
\bibitem [{\citenamefont {Macio{\l}ek}\ and\ \citenamefont
  {Dietrich}(2018)}]{MD-18}%
  \BibitemOpen
  \bibfield  {author} {\bibinfo {author} {\bibfnamefont {A.}~\bibnamefont
  {Macio{\l}ek}}\ and\ \bibinfo {author} {\bibfnamefont {S.}~\bibnamefont
  {Dietrich}},\ }\bibfield  {title} {\bibinfo {title} {{Collective behavior of
  colloids due to critical Casimir interactions}},\ }\href
  {https://doi.org/10.1103/RevModPhys.90.045001} {\bibfield  {journal}
  {\bibinfo  {journal} {\RMP}\ }\textbf {\bibinfo {volume} {90}},\ \bibinfo
  {eid} {045001} (\bibinfo {year} {2018})},\ \Eprint
  {https://arxiv.org/abs/1712.06678} {arXiv:1712.06678 [cond-mat.soft]}
  \BibitemShut {NoStop}%
\bibitem [{\citenamefont {{Pelissetto}}\ and\ \citenamefont
  {{Vicari}}(2002)}]{PV-02}%
  \BibitemOpen
  \bibfield  {author} {\bibinfo {author} {\bibfnamefont {A.}~\bibnamefont
  {{Pelissetto}}}\ and\ \bibinfo {author} {\bibfnamefont {E.}~\bibnamefont
  {{Vicari}}},\ }\bibfield  {title} {\bibinfo {title} {{Critical phenomena and
  renormalization-group theory}},\ }\href
  {https://doi.org/10.1016/S0370-1573(02)00219-3} {\bibfield  {journal}
  {\bibinfo  {journal} {\physrep}\ }\textbf {\bibinfo {volume} {368}},\
  \bibinfo {pages} {549} (\bibinfo {year} {2002})},\ \Eprint
  {https://arxiv.org/abs/cond-mat/0012164} {cond-mat/0012164} \BibitemShut
  {NoStop}%
\bibitem [{\citenamefont {Arnold}\ and\ \citenamefont {Pappas}(2000)}]{AP-00}%
  \BibitemOpen
  \bibfield  {author} {\bibinfo {author} {\bibfnamefont {C.~S.}\ \bibnamefont
  {Arnold}}\ and\ \bibinfo {author} {\bibfnamefont {D.~P.}\ \bibnamefont
  {Pappas}},\ }\bibfield  {title} {\bibinfo {title} {Gd(0001): A semi-infinite
  three-dimensional heisenberg ferromagnet with ordinary surface transition},\
  }\href {https://doi.org/10.1103/PhysRevLett.85.5202} {\bibfield  {journal}
  {\bibinfo  {journal} {\prl}\ }\textbf {\bibinfo {volume} {85}},\ \bibinfo
  {pages} {5202} (\bibinfo {year} {2000})}\BibitemShut {NoStop}%
\bibitem [{\citenamefont {Mermin}\ and\ \citenamefont
  {Wagner}(1966{\natexlab{a}})}]{MW-66}%
  \BibitemOpen
  \bibfield  {author} {\bibinfo {author} {\bibfnamefont {N.~D.}\ \bibnamefont
  {Mermin}}\ and\ \bibinfo {author} {\bibfnamefont {H.}~\bibnamefont
  {Wagner}},\ }\bibfield  {title} {\bibinfo {title} {{Absence of Ferromagnetism
  or Antiferromagnetism in One- or Two-Dimensional Isotropic Heisenberg
  Models}},\ }\href {https://doi.org/10.1103/PhysRevLett.17.1133} {\bibfield
  {journal} {\bibinfo  {journal} {\PRL}\ }\textbf {\bibinfo {volume} {17}},\
  \bibinfo {pages} {1133} (\bibinfo {year} {1966}{\natexlab{a}})}\BibitemShut
  {NoStop}%
\bibitem [{\citenamefont {Mermin}\ and\ \citenamefont
  {Wagner}(1966{\natexlab{b}})}]{MW-66_erratum}%
  \BibitemOpen
  \bibfield  {author} {\bibinfo {author} {\bibfnamefont {N.~D.}\ \bibnamefont
  {Mermin}}\ and\ \bibinfo {author} {\bibfnamefont {H.}~\bibnamefont
  {Wagner}},\ }\bibfield  {title} {\bibinfo {title} {{Erratum: Absence of
  Ferromagnetism or Antiferromagnetism in One- or Two-Dimensional Isotropic
  Heisenberg Models}},\ }\href
  {https://doi.org/https://doi.org/10.1103/PhysRevLett.17.1307} {\bibfield
  {journal} {\bibinfo  {journal} {\PRL}\ }\textbf {\bibinfo {volume} {17}},\
  \bibinfo {pages} {1307} (\bibinfo {year} {1966}{\natexlab{b}})}\BibitemShut
  {NoStop}%
\bibitem [{\citenamefont {Hohenberg}(1967)}]{Hohenberg-67}%
  \BibitemOpen
  \bibfield  {author} {\bibinfo {author} {\bibfnamefont {P.~C.}\ \bibnamefont
  {Hohenberg}},\ }\bibfield  {title} {\bibinfo {title} {{Existence of
  Long-Range Order in One and Two Dimensions}},\ }\href
  {https://doi.org/10.1103/PhysRev.158.383} {\bibfield  {journal} {\bibinfo
  {journal} {\PR}\ }\textbf {\bibinfo {volume} {158}},\ \bibinfo {pages} {383}
  (\bibinfo {year} {1967})}\BibitemShut {NoStop}%
\bibitem [{\citenamefont {Halperin}(2019)}]{Halperin-19}%
  \BibitemOpen
  \bibfield  {author} {\bibinfo {author} {\bibfnamefont {B.~I.}\ \bibnamefont
  {Halperin}},\ }\bibfield  {title} {\bibinfo {title} {{On the
  Hohenberg-Mermin-Wagner Theorem and Its Limitations}},\ }\href
  {https://doi.org/10.1007/s10955-018-2202-y} {\bibfield  {journal} {\bibinfo
  {journal} {\JSP}\ }\textbf {\bibinfo {volume} {175}},\ \bibinfo {pages} {521}
  (\bibinfo {year} {2019})},\ \Eprint {https://arxiv.org/abs/1812.00220}
  {arXiv:1812.00220 [cond-mat.stat-mech]} \BibitemShut {NoStop}%
\bibitem [{\citenamefont {Krech}(2000)}]{Krech-00}%
  \BibitemOpen
  \bibfield  {author} {\bibinfo {author} {\bibfnamefont {M.}~\bibnamefont
  {Krech}},\ }\bibfield  {title} {\bibinfo {title} {Surface scaling behavior of
  isotropic heisenberg systems: Critical exponents, structure factor, and
  profiles},\ }\href {https://doi.org/10.1103/PhysRevB.62.6360} {\bibfield
  {journal} {\bibinfo  {journal} {\prb}\ }\textbf {\bibinfo {volume} {62}},\
  \bibinfo {pages} {6360} (\bibinfo {year} {2000})},\ \Eprint
  {https://arxiv.org/abs/cond-mat/0006448} {arXiv:cond-mat/0006448
  [cond-mat.stat-mech]} \BibitemShut {NoStop}%
\bibitem [{\citenamefont {{Deng}}\ \emph {et~al.}(2005)\citenamefont {{Deng}},
  \citenamefont {{Bl{\"o}te}},\ and\ \citenamefont {{Nightingale}}}]{DBN-05}%
  \BibitemOpen
  \bibfield  {author} {\bibinfo {author} {\bibfnamefont {Y.}~\bibnamefont
  {{Deng}}}, \bibinfo {author} {\bibfnamefont {H.~W.~J.}\ \bibnamefont
  {{Bl{\"o}te}}},\ and\ \bibinfo {author} {\bibfnamefont {M.~P.}\ \bibnamefont
  {{Nightingale}}},\ }\bibfield  {title} {\bibinfo {title} {{Surface and bulk
  transitions in three-dimensional O(n) models}},\ }\href
  {https://doi.org/10.1103/PhysRevE.72.016128} {\bibfield  {journal} {\bibinfo
  {journal} {\pre}\ }\textbf {\bibinfo {volume} {72}},\ \bibinfo {eid} {016128}
  (\bibinfo {year} {2005})},\ \Eprint {https://arxiv.org/abs/cond-mat/0504173}
  {cond-mat/0504173} \BibitemShut {NoStop}%
\bibitem [{\citenamefont {{Suzuki}}\ and\ \citenamefont
  {{Sato}}(2012)}]{SS-12}%
  \BibitemOpen
  \bibfield  {author} {\bibinfo {author} {\bibfnamefont {T.}~\bibnamefont
  {{Suzuki}}}\ and\ \bibinfo {author} {\bibfnamefont {M.}~\bibnamefont
  {{Sato}}},\ }\bibfield  {title} {\bibinfo {title} {{Gapless edge states and
  their stability in two-dimensional quantum magnets}},\ }\href
  {https://doi.org/10.1103/PhysRevB.86.224411} {\bibfield  {journal} {\bibinfo
  {journal} {\prb}\ }\textbf {\bibinfo {volume} {86}},\ \bibinfo {eid} {224411}
  (\bibinfo {year} {2012})},\ \Eprint {https://arxiv.org/abs/1209.3097}
  {arXiv:1209.3097 [cond-mat.str-el]} \BibitemShut {NoStop}%
\bibitem [{\citenamefont {Zhang}\ and\ \citenamefont {Wang}(2017)}]{ZW-17}%
  \BibitemOpen
  \bibfield  {author} {\bibinfo {author} {\bibfnamefont {L.}~\bibnamefont
  {Zhang}}\ and\ \bibinfo {author} {\bibfnamefont {F.}~\bibnamefont {Wang}},\
  }\bibfield  {title} {\bibinfo {title} {{Unconventional Surface Critical
  Behavior Induced by a Quantum Phase Transition from the Two-Dimensional
  Affleck-Kennedy-Lieb-Tasaki Phase to a N{\'e}el-Ordered Phase}},\ }\href
  {https://doi.org/10.1103/PhysRevLett.118.087201} {\bibfield  {journal}
  {\bibinfo  {journal} {\PRL}\ }\textbf {\bibinfo {volume} {118}},\ \bibinfo
  {eid} {087201} (\bibinfo {year} {2017})},\ \Eprint
  {https://arxiv.org/abs/1611.06477} {arXiv:1611.06477 [cond-mat.str-el]}
  \BibitemShut {NoStop}%
\bibitem [{\citenamefont {Ding}\ \emph {et~al.}(2018)\citenamefont {Ding},
  \citenamefont {Zhang},\ and\ \citenamefont {Guo}}]{DZG-18}%
  \BibitemOpen
  \bibfield  {author} {\bibinfo {author} {\bibfnamefont {C.}~\bibnamefont
  {Ding}}, \bibinfo {author} {\bibfnamefont {L.}~\bibnamefont {Zhang}},\ and\
  \bibinfo {author} {\bibfnamefont {W.}~\bibnamefont {Guo}},\ }\bibfield
  {title} {\bibinfo {title} {{Engineering Surface Critical Behavior of $(2 +1
  )$-Dimensional O(3) Quantum Critical Points}},\ }\href
  {https://doi.org/10.1103/PhysRevLett.120.235701} {\bibfield  {journal}
  {\bibinfo  {journal} {\PRL}\ }\textbf {\bibinfo {volume} {120}},\ \bibinfo
  {eid} {235701} (\bibinfo {year} {2018})},\ \Eprint
  {https://arxiv.org/abs/1801.10035} {arXiv:1801.10035 [cond-mat.str-el]}
  \BibitemShut {NoStop}%
\bibitem [{\citenamefont {Weber}\ \emph {et~al.}(2018)\citenamefont {Weber},
  \citenamefont {Parisen~Toldin},\ and\ \citenamefont {Wessel}}]{WPTW-18}%
  \BibitemOpen
  \bibfield  {author} {\bibinfo {author} {\bibfnamefont {L.}~\bibnamefont
  {Weber}}, \bibinfo {author} {\bibfnamefont {F.}~\bibnamefont
  {Parisen~Toldin}},\ and\ \bibinfo {author} {\bibfnamefont {S.}~\bibnamefont
  {Wessel}},\ }\bibfield  {title} {\bibinfo {title} {Nonordinary edge
  criticality of two-dimensional quantum critical magnets},\ }\href
  {https://doi.org/10.1103/PhysRevB.98.140403} {\bibfield  {journal} {\bibinfo
  {journal} {\prb}\ }\textbf {\bibinfo {volume} {98}},\ \bibinfo {eid}
  {140403(R)} (\bibinfo {year} {2018})},\ \Eprint
  {https://arxiv.org/abs/1804.06820} {arXiv:1804.06820 [cond-mat.str-el]}
  \BibitemShut {NoStop}%
\bibitem [{\citenamefont {Jian}\ \emph {et~al.}(2021)\citenamefont {Jian},
  \citenamefont {Xu}, \citenamefont {Wu},\ and\ \citenamefont {Xu}}]{JXWX-20}%
  \BibitemOpen
  \bibfield  {author} {\bibinfo {author} {\bibfnamefont {C.-M.}\ \bibnamefont
  {Jian}}, \bibinfo {author} {\bibfnamefont {Y.}~\bibnamefont {Xu}}, \bibinfo
  {author} {\bibfnamefont {X.-C.}\ \bibnamefont {Wu}},\ and\ \bibinfo {author}
  {\bibfnamefont {C.}~\bibnamefont {Xu}},\ }\bibfield  {title} {\bibinfo
  {title} {{Continuous N{\'e}el-VBS quantum phase transition in non-local
  one-dimensional systems with SO(3) symmetry}},\ }\href
  {https://doi.org/10.21468/SciPostPhys.10.2.033} {\bibfield  {journal}
  {\bibinfo  {journal} {SciPost Phys.}\ }\textbf {\bibinfo {volume} {10}},\
  \bibinfo {eid} {033} (\bibinfo {year} {2021})},\ \Eprint
  {https://arxiv.org/abs/2004.07852} {arXiv:2004.07852 [cond-mat.str-el]}
  \BibitemShut {NoStop}%
\bibitem [{\citenamefont {Weber}\ and\ \citenamefont {Wessel}(2019)}]{WW-19}%
  \BibitemOpen
  \bibfield  {author} {\bibinfo {author} {\bibfnamefont {L.}~\bibnamefont
  {Weber}}\ and\ \bibinfo {author} {\bibfnamefont {S.}~\bibnamefont {Wessel}},\
  }\bibfield  {title} {\bibinfo {title} {Nonordinary criticality at the edges
  of planar spin-1 heisenberg antiferromagnets},\ }\href
  {https://doi.org/10.1103/PhysRevB.100.054437} {\bibfield  {journal} {\bibinfo
   {journal} {\prb}\ }\textbf {\bibinfo {volume} {100}},\ \bibinfo {eid}
  {054437} (\bibinfo {year} {2019})},\ \Eprint
  {https://arxiv.org/abs/1906.07051} {arXiv:1906.07051 [cond-mat.str-el]}
  \BibitemShut {NoStop}%
\bibitem [{\citenamefont {Weber}\ and\ \citenamefont {Wessel}(2021)}]{WW-20}%
  \BibitemOpen
  \bibfield  {author} {\bibinfo {author} {\bibfnamefont {L.}~\bibnamefont
  {Weber}}\ and\ \bibinfo {author} {\bibfnamefont {S.}~\bibnamefont {Wessel}},\
  }\bibfield  {title} {\bibinfo {title} {Spin versus bond correlations along
  dangling edges of quantum critical magnets},\ }\href
  {https://doi.org/10.1103/PhysRevB.103.L020406} {\bibfield  {journal}
  {\bibinfo  {journal} {\PRB}\ }\textbf {\bibinfo {volume} {103}},\ \bibinfo
  {eid} {arXiv:2010.15691} (\bibinfo {year} {2021})},\ \Eprint
  {https://arxiv.org/abs/2010.15691} {arXiv:2010.15691 [cond-mat.str-el]}
  \BibitemShut {NoStop}%
\bibitem [{\citenamefont {Zhu}\ \emph {et~al.}(2021)\citenamefont {Zhu},
  \citenamefont {Ding}, \citenamefont {Zhang},\ and\ \citenamefont
  {Guo}}]{ZDZG-20}%
  \BibitemOpen
  \bibfield  {author} {\bibinfo {author} {\bibfnamefont {W.}~\bibnamefont
  {Zhu}}, \bibinfo {author} {\bibfnamefont {C.}~\bibnamefont {Ding}}, \bibinfo
  {author} {\bibfnamefont {L.}~\bibnamefont {Zhang}},\ and\ \bibinfo {author}
  {\bibfnamefont {W.}~\bibnamefont {Guo}},\ }\bibfield  {title} {\bibinfo
  {title} {{Surface critical behavior of coupled Haldane chains}},\ }\href
  {https://doi.org/10.1103/PhysRevB.103.024412} {\bibfield  {journal} {\bibinfo
   {journal} {\prb}\ }\textbf {\bibinfo {volume} {103}},\ \bibinfo {eid}
  {024412} (\bibinfo {year} {2021})},\ \Eprint
  {https://arxiv.org/abs/2010.10920} {arXiv:2010.10920 [cond-mat.str-el]}
  \BibitemShut {NoStop}%
\bibitem [{\citenamefont {{Sachdev}}(2011)}]{Sachdev-book}%
  \BibitemOpen
  \bibfield  {author} {\bibinfo {author} {\bibfnamefont {S.}~\bibnamefont
  {{Sachdev}}},\ }\href@noop {} {\emph {\bibinfo {title} {{Quantum Phase
  Transitions}}}}\ (\bibinfo  {publisher} {{Cambridge University Press}},\
  \bibinfo {address} {{Cambridge, UK}},\ \bibinfo {year} {2011})\BibitemShut
  {NoStop}%
\bibitem [{\citenamefont {Metlitski}(2022)}]{Metlitski-20}%
  \BibitemOpen
  \bibfield  {author} {\bibinfo {author} {\bibfnamefont {M.~A.}\ \bibnamefont
  {Metlitski}},\ }\bibfield  {title} {\bibinfo {title} {{Boundary criticality
  of the O(N) model in d = 3 critically revisited}},\ }\href
  {https://doi.org/10.21468/SciPostPhys.12.4.131} {\bibfield  {journal}
  {\bibinfo  {journal} {SciPost Phys.}\ }\textbf {\bibinfo {volume} {12}},\
  \bibinfo {pages} {131} (\bibinfo {year} {2022})},\ \Eprint
  {https://arxiv.org/abs/2009.05119} {arXiv:2009.05119 [cond-mat.str-el]}
  \BibitemShut {NoStop}%
\bibitem [{\citenamefont {{Diehl}}(1994)}]{Diehl-94}%
  \BibitemOpen
  \bibfield  {author} {\bibinfo {author} {\bibfnamefont {H.~W.}\ \bibnamefont
  {{Diehl}}},\ }\bibfield  {title} {\bibinfo {title} {{Critical adsorption of
  fluids and the equivalence of extraordinary to normal surface transitions}},\
  }\href {https://doi.org/10.1002/bbpc.19940980344} {\bibfield  {journal}
  {\bibinfo  {journal} {Ber. Bunsenges. Phys. Chem.}\ }\textbf {\bibinfo
  {volume} {98}},\ \bibinfo {pages} {466} (\bibinfo {year} {1994})}\BibitemShut
  {NoStop}%
\bibitem [{\citenamefont {{Burkhardt}}\ and\ \citenamefont
  {{Diehl}}(1994)}]{BD-94}%
  \BibitemOpen
  \bibfield  {author} {\bibinfo {author} {\bibfnamefont {T.~W.}\ \bibnamefont
  {{Burkhardt}}}\ and\ \bibinfo {author} {\bibfnamefont {H.~W.}\ \bibnamefont
  {{Diehl}}},\ }\bibfield  {title} {\bibinfo {title} {{Ordinary, extraordinary,
  and normal surface transitions: Extraordinary-normal equivalence and simple
  explanation of $|T-T_c|^{2-\alpha}$ singularities}},\ }\href
  {https://doi.org/10.1103/PhysRevB.50.3894} {\bibfield  {journal} {\bibinfo
  {journal} {\prb}\ }\textbf {\bibinfo {volume} {50}},\ \bibinfo {pages} {3894}
  (\bibinfo {year} {1994})},\ \Eprint {https://arxiv.org/abs/cond-mat/9402077}
  {cond-mat/9402077} \BibitemShut {NoStop}%
\bibitem [{\citenamefont {{Campostrini}}\ \emph {et~al.}(2002)\citenamefont
  {{Campostrini}}, \citenamefont {{Hasenbusch}}, \citenamefont {{Pelissetto}},
  \citenamefont {{Rossi}},\ and\ \citenamefont {{Vicari}}}]{CHPRV-02}%
  \BibitemOpen
  \bibfield  {author} {\bibinfo {author} {\bibfnamefont {M.}~\bibnamefont
  {{Campostrini}}}, \bibinfo {author} {\bibfnamefont {M.}~\bibnamefont
  {{Hasenbusch}}}, \bibinfo {author} {\bibfnamefont {A.}~\bibnamefont
  {{Pelissetto}}}, \bibinfo {author} {\bibfnamefont {P.}~\bibnamefont
  {{Rossi}}},\ and\ \bibinfo {author} {\bibfnamefont {E.}~\bibnamefont
  {{Vicari}}},\ }\bibfield  {title} {\bibinfo {title} {{Critical exponents and
  equation of state of the three-dimensional Heisenberg universality class}},\
  }\href {https://doi.org/10.1103/PhysRevB.65.144520} {\bibfield  {journal}
  {\bibinfo  {journal} {\prb}\ }\textbf {\bibinfo {volume} {65}},\ \bibinfo
  {eid} {144520} (\bibinfo {year} {2002})},\ \Eprint
  {https://arxiv.org/abs/cond-mat/0110336} {cond-mat/0110336} \BibitemShut
  {NoStop}%
\bibitem [{\citenamefont {Hasenbusch}(2020)}]{Hasenbusch-20}%
  \BibitemOpen
  \bibfield  {author} {\bibinfo {author} {\bibfnamefont {M.}~\bibnamefont
  {Hasenbusch}},\ }\bibfield  {title} {\bibinfo {title} {Monte carlo study of a
  generalized icosahedral model on the simple cubic lattice},\ }\href
  {https://doi.org/10.1103/PhysRevB.102.024406} {\bibfield  {journal} {\bibinfo
   {journal} {\prb}\ }\textbf {\bibinfo {volume} {102}},\ \bibinfo {eid}
  {024406} (\bibinfo {year} {2020})},\ \Eprint
  {https://arxiv.org/abs/2005.04448} {arXiv:2005.04448 [cond-mat.stat-mech]}
  \BibitemShut {NoStop}%
\bibitem [{\citenamefont {{Newman}}\ and\ \citenamefont
  {{Riedel}}(1984)}]{NR-84}%
  \BibitemOpen
  \bibfield  {author} {\bibinfo {author} {\bibfnamefont {K.~E.}\ \bibnamefont
  {{Newman}}}\ and\ \bibinfo {author} {\bibfnamefont {E.~K.}\ \bibnamefont
  {{Riedel}}},\ }\bibfield  {title} {\bibinfo {title} {{Critical exponents by
  the scaling-field method: The isotropic N-vector model in three
  dimensions}},\ }\href {https://doi.org/10.1103/PhysRevB.30.6615} {\bibfield
  {journal} {\bibinfo  {journal} {\prb}\ }\textbf {\bibinfo {volume} {30}},\
  \bibinfo {pages} {6615} (\bibinfo {year} {1984})}\BibitemShut {NoStop}%
\bibitem [{\citenamefont {{Hasenbusch}}(2009)}]{Hasenbusch-09b}%
  \BibitemOpen
  \bibfield  {author} {\bibinfo {author} {\bibfnamefont {M.}~\bibnamefont
  {{Hasenbusch}}},\ }\bibfield  {title} {\bibinfo {title} {{The thermodynamic
  Casimir effect in the neighbourhood of the {$\lambda$}-transition: a Monte
  Carlo study of an improved three-dimensional lattice model}},\ }\href
  {https://doi.org/10.1088/1742-5468/2009/07/P07031} {\bibfield  {journal}
  {\bibinfo  {journal} {\JSTAT}\ } (\bibinfo {year} 2009) {\bibinfo {volume} {P07031}},\
  \bibinfo {pages} {{}} } \Eprint
  {https://arxiv.org/abs/0905.2096} {arXiv:0905.2096 [cond-mat.stat-mech]}
  \BibitemShut {NoStop}%
\bibitem [{\citenamefont {{Hasenbusch}}(2010{\natexlab{a}})}]{Hasenbusch-10c}%
  \BibitemOpen
  \bibfield  {author} {\bibinfo {author} {\bibfnamefont {M.}~\bibnamefont
  {{Hasenbusch}}},\ }\bibfield  {title} {\bibinfo {title} {{Thermodynamic
  Casimir effect for films in the three-dimensional Ising universality class:
  Symmetry-breaking boundary conditions}},\ }\href
  {https://doi.org/10.1103/PhysRevB.82.104425} {\bibfield  {journal} {\bibinfo
  {journal} {\prb}\ }\textbf {\bibinfo {volume} {82}},\ \bibinfo {eid} {104425}
  (\bibinfo {year} {2010}{\natexlab{a}})},\ \Eprint
  {https://arxiv.org/abs/1005.4749} {arXiv:1005.4749 [cond-mat.stat-mech]}
  \BibitemShut {NoStop}%
\bibitem [{\citenamefont {{Parisen Toldin}}\ and\ \citenamefont
  {{Dietrich}}(2010)}]{PTD-10}%
  \BibitemOpen
  \bibfield  {author} {\bibinfo {author} {\bibfnamefont {F.}~\bibnamefont
  {{Parisen Toldin}}}\ and\ \bibinfo {author} {\bibfnamefont {S.}~\bibnamefont
  {{Dietrich}}},\ }\bibfield  {title} {\bibinfo {title} {{Critical Casimir
  forces and adsorption profiles in the presence of a chemically structured
  substrate}},\ }\href {https://doi.org/10.1088/1742-5468/2010/11/P11003}
  {\bibfield  {journal} {\bibinfo  {journal} {\JSTAT}\ } (\bibinfo {year} 2010) {\bibinfo
  {volume} {P11003}},\ \bibinfo {pages} {{}} }
  \Eprint {https://arxiv.org/abs/1007.3913} {arXiv:1007.3913
  [cond-mat.stat-mech]} \BibitemShut {NoStop}%
\bibitem [{\citenamefont {{Hasenbusch}}(2011)}]{Hasenbusch-11}%
  \BibitemOpen
  \bibfield  {author} {\bibinfo {author} {\bibfnamefont {M.}~\bibnamefont
  {{Hasenbusch}}},\ }\bibfield  {title} {\bibinfo {title} {{Thermodynamic
  Casimir force: A Monte Carlo study of the crossover between the ordinary and
  the normal surface universality class}},\ }\href
  {https://doi.org/10.1103/PhysRevB.83.134425} {\bibfield  {journal} {\bibinfo
  {journal} {\prb}\ }\textbf {\bibinfo {volume} {83}},\ \bibinfo {eid} {134425}
  (\bibinfo {year} {2011})},\ \Eprint {https://arxiv.org/abs/1012.4986}
  {arXiv:1012.4986 [cond-mat.stat-mech]} \BibitemShut {NoStop}%
\bibitem [{\citenamefont {Hasenbusch}(2011)}]{Hasenbusch-11b}%
  \BibitemOpen
  \bibfield  {author} {\bibinfo {author} {\bibfnamefont {M.}~\bibnamefont
  {Hasenbusch}},\ }\bibfield  {title} {\bibinfo {title} {{Monte Carlo study of
  surface critical phenomena: The special point}},\ }\href
  {https://doi.org/10.1103/PhysRevB.84.134405} {\bibfield  {journal} {\bibinfo
  {journal} {\prb}\ }\textbf {\bibinfo {volume} {84}},\ \bibinfo {eid} {134405}
  (\bibinfo {year} {2011})},\ \Eprint {https://arxiv.org/abs/1108.2425}
  {arXiv:1108.2425 [cond-mat.stat-mech]} \BibitemShut {NoStop}%
\bibitem [{\citenamefont {{Hasenbusch}}(2012)}]{Hasenbusch-12}%
  \BibitemOpen
  \bibfield  {author} {\bibinfo {author} {\bibfnamefont {M.}~\bibnamefont
  {{Hasenbusch}}},\ }\bibfield  {title} {\bibinfo {title} {{Thermodynamic
  Casimir effect: Universality and corrections to scaling}},\ }\href
  {https://doi.org/10.1103/PhysRevB.85.174421} {\bibfield  {journal} {\bibinfo
  {journal} {\prb}\ }\textbf {\bibinfo {volume} {85}},\ \bibinfo {eid} {174421}
  (\bibinfo {year} {2012})},\ \Eprint {https://arxiv.org/abs/1202.6206}
  {arXiv:1202.6206 [cond-mat.stat-mech]} \BibitemShut {NoStop}%
\bibitem [{\citenamefont {{Parisen Toldin}}\ \emph {et~al.}(2013)\citenamefont
  {{Parisen Toldin}}, \citenamefont {{Tr{\"o}ndle}},\ and\ \citenamefont
  {{Dietrich}}}]{PTTD-13}%
  \BibitemOpen
  \bibfield  {author} {\bibinfo {author} {\bibfnamefont {F.}~\bibnamefont
  {{Parisen Toldin}}}, \bibinfo {author} {\bibfnamefont {M.}~\bibnamefont
  {{Tr{\"o}ndle}}},\ and\ \bibinfo {author} {\bibfnamefont {S.}~\bibnamefont
  {{Dietrich}}},\ }\bibfield  {title} {\bibinfo {title} {{Critical Casimir
  forces between homogeneous and chemically striped surfaces}},\ }\href
  {https://doi.org/10.1103/PhysRevE.88.052110} {\bibfield  {journal} {\bibinfo
  {journal} {\pre}\ }\textbf {\bibinfo {volume} {88}},\ \bibinfo {eid} {052110}
  (\bibinfo {year} {2013})},\ \Eprint {https://arxiv.org/abs/1303.6104}
  {arXiv:1303.6104 [cond-mat.stat-mech]} \BibitemShut {NoStop}%
\bibitem [{\citenamefont {{Parisen Toldin}}(2015)}]{PT-13}%
  \BibitemOpen
  \bibfield  {author} {\bibinfo {author} {\bibfnamefont {F.}~\bibnamefont
  {{Parisen Toldin}}},\ }\bibfield  {title} {\bibinfo {title} {{Critical
  Casimir force in the presence of random local adsorption preference}},\
  }\href {https://doi.org/10.1103/PhysRevE.91.032105} {\bibfield  {journal}
  {\bibinfo  {journal} {\PRE}\ }\textbf {\bibinfo {volume} {91}},\ \bibinfo
  {pages} {032105} (\bibinfo {year} {2015})},\ \Eprint
  {https://arxiv.org/abs/1308.5220} {arXiv:1308.5220 [cond-mat.stat-mech]}
  \BibitemShut {NoStop}%
\bibitem [{\citenamefont {{Parisen Toldin}}\ \emph
  {et~al.}(2015{\natexlab{a}})\citenamefont {{Parisen Toldin}}, \citenamefont
  {{Tr{\"o}ndle}},\ and\ \citenamefont {{Dietrich}}}]{PTTD-14}%
  \BibitemOpen
  \bibfield  {author} {\bibinfo {author} {\bibfnamefont {F.}~\bibnamefont
  {{Parisen Toldin}}}, \bibinfo {author} {\bibfnamefont {M.}~\bibnamefont
  {{Tr{\"o}ndle}}},\ and\ \bibinfo {author} {\bibfnamefont {S.}~\bibnamefont
  {{Dietrich}}},\ }\bibfield  {title} {\bibinfo {title} {{Line contribution to
  the critical Casimir force between a homogeneous and a chemically stepped
  surface}},\ }\href {https://doi.org/10.1088/0953-8984/27/21/214010}
  {\bibfield  {journal} {\bibinfo  {journal} {\JPCM}\ }\textbf {\bibinfo
  {volume} {27}},\ \bibinfo {eid} {214010} (\bibinfo {year}
  {2015}{\natexlab{a}})},\ \Eprint {https://arxiv.org/abs/1409.5536}
  {arXiv:1409.5536 [cond-mat.stat-mech]} \BibitemShut {NoStop}%
\bibitem [{\citenamefont {{Parisen Toldin}}\ \emph {et~al.}(2017)\citenamefont
  {{Parisen Toldin}}, \citenamefont {{Assaad}},\ and\ \citenamefont
  {{Wessel}}}]{PTAW-17}%
  \BibitemOpen
  \bibfield  {author} {\bibinfo {author} {\bibfnamefont {F.}~\bibnamefont
  {{Parisen Toldin}}}, \bibinfo {author} {\bibfnamefont {F.~F.}\ \bibnamefont
  {{Assaad}}},\ and\ \bibinfo {author} {\bibfnamefont {S.}~\bibnamefont
  {{Wessel}}},\ }\bibfield  {title} {\bibinfo {title} {{Critical behavior in
  the presence of an order-parameter pinning field}},\ }\href
  {https://doi.org/10.1103/PhysRevB.95.014401} {\bibfield  {journal} {\bibinfo
  {journal} {\prb}\ }\textbf {\bibinfo {volume} {95}},\ \bibinfo {eid} {014401}
  (\bibinfo {year} {2017})},\ \Eprint {https://arxiv.org/abs/1607.04270}
  {arXiv:1607.04270 [cond-mat.stat-mech]} \BibitemShut {NoStop}%
\bibitem [{\citenamefont {{Wolff}}(1989)}]{Wolff-89}%
  \BibitemOpen
  \bibfield  {author} {\bibinfo {author} {\bibfnamefont {U.}~\bibnamefont
  {{Wolff}}},\ }\bibfield  {title} {\bibinfo {title} {{Collective Monte Carlo
  updating for spin systems}},\ }\href
  {https://doi.org/10.1103/PhysRevLett.62.361} {\bibfield  {journal} {\bibinfo
  {journal} {\PRL}\ }\textbf {\bibinfo {volume} {62}},\ \bibinfo {pages} {361}
  (\bibinfo {year} {1989})}\BibitemShut {NoStop}%
\bibitem [{See Supplemental Material for technical details on the Monte Carlo
  simulations, additional finite-size scaling analysis, and a comparison with
  available field-theory results.()}]{SM}%
  \BibitemOpen
  See Supplemental Material for technical details on the Monte Carlo
  simulations, additional finite-size scaling analysis, and a comparison with
  available field-theory results\ \href@noop {} {}\BibitemShut {NoStop}%
\bibitem [{\citenamefont {{Parisen Toldin}}(2023)}]{PT-23}%
  \BibitemOpen
  \bibfield  {author} {\bibinfo {author} {\bibfnamefont {F.}~\bibnamefont
  {{Parisen Toldin}}},\ }\bibfield  {title} {\bibinfo {title} {{The ordinary
  surface universality class of the three-dimensional O ($N$) model}},\ }\href
  {https://doi.org/10.1103/PhysRevB.108.L020404} {\bibfield  {journal}
  {\bibinfo  {journal} {\prb}\ }\textbf {\bibinfo {volume} {108}},\ \bibinfo
  {eid} {L020404} (\bibinfo {year} {2023})},\ \Eprint
  {https://arxiv.org/abs/2303.16683} {arXiv:2303.16683 [cond-mat.stat-mech]}
  \BibitemShut {NoStop}%
\bibitem [{\citenamefont {{Privman}}(1990)}]{Privman-90}%
  \BibitemOpen
  \bibfield  {author} {\bibinfo {author} {\bibfnamefont {V.}~\bibnamefont
  {{Privman}}},\ }\bibfield  {title} {\bibinfo {title} {{Finite-Size Scaling
  Theory}},\ }in\ \href
  {http://www.worldscientific.com/worldscibooks/10.1142/1011} {\emph {\bibinfo
  {booktitle} {Finite Size Scaling and Numerical Simulation of Statistical
  Systems}}},\ \bibinfo {editor} {edited by\ \bibinfo {editor} {\bibfnamefont
  {V.}~\bibnamefont {{Privman}}}}\ (\bibinfo  {publisher} {World Scientific},\
  \bibinfo {address} {Singapore},\ \bibinfo {year} {1990})\ p.~\bibinfo {pages}
  {1}\BibitemShut {NoStop}%
\bibitem [{\citenamefont {Caracciolo}\ and\ \citenamefont
  {Sokal}(1986)}]{CS-86}%
  \BibitemOpen
  \bibfield  {author} {\bibinfo {author} {\bibfnamefont {S.}~\bibnamefont
  {Caracciolo}}\ and\ \bibinfo {author} {\bibfnamefont {A.~D.}\ \bibnamefont
  {Sokal}},\ }\bibfield  {title} {\bibinfo {title} {{Dynamic critical exponent
  of some Monte Carlo algorithms for the self-avoiding walk}},\ }\href
  {https://doi.org/10.1088/0305-4470/19/13/008} {\bibfield  {journal} {\bibinfo
   {journal} {\JPAOLD}\ }\textbf {\bibinfo {volume} {19}},\ \bibinfo {pages}
  {L797} (\bibinfo {year} {1986})}\BibitemShut {NoStop}%
\bibitem [{\citenamefont {Madras}\ and\ \citenamefont {Sokal}(1988)}]{MS-88}%
  \BibitemOpen
  \bibfield  {author} {\bibinfo {author} {\bibfnamefont {N.}~\bibnamefont
  {Madras}}\ and\ \bibinfo {author} {\bibfnamefont {A.~D.}\ \bibnamefont
  {Sokal}},\ }\bibfield  {title} {\bibinfo {title} {{The pivot algorithm: A
  highly efficient Monte Carlo method for the self-avoiding walk}},\ }\href
  {https://doi.org/10.1007/BF01022990} {\bibfield  {journal} {\bibinfo
  {journal} {\JSP}\ }\textbf {\bibinfo {volume} {50}},\ \bibinfo {pages} {109}
  (\bibinfo {year} {1988})}\BibitemShut {NoStop}%
\bibitem [{\citenamefont {Sokal}(1997)}]{Sokal_lecture}%
  \BibitemOpen
  \bibfield  {author} {\bibinfo {author} {\bibfnamefont {A.}~\bibnamefont
  {Sokal}},\ }\bibinfo {title} {{Monte Carlo Methods in Statistical Mechanics:
  Foundations and New Algorithms}},\ in\ \href
  {https://doi.org/10.1007/978-1-4899-0319-8_6} {\emph {\bibinfo {booktitle}
  {Functional Integration: Basics and Applications}}},\ \bibinfo {editor}
  {edited by\ \bibinfo {editor} {\bibfnamefont {C.}~\bibnamefont
  {DeWitt-Morette}}, \bibinfo {editor} {\bibfnamefont {P.}~\bibnamefont
  {Cartier}},\ and\ \bibinfo {editor} {\bibfnamefont {A.}~\bibnamefont
  {Folacci}}}\ (\bibinfo  {publisher} {Springer US},\ \bibinfo {address}
  {Boston, MA},\ \bibinfo {year} {1997})\ pp.\ \bibinfo {pages}
  {131--192}\BibitemShut {NoStop}%
\bibitem [{\citenamefont {Young}(2015)}]{Young_notes}%
  \BibitemOpen
  \bibfield  {author} {\bibinfo {author} {\bibfnamefont {A.~P.}\ \bibnamefont
  {Young}},\ }\href {https://doi.org/10.1007/978-3-319-19051-8} {\emph
  {\bibinfo {title} {{Everything You Wanted to Know About Data Analysis and
  Fitting but Were Afraid to Ask}}}},\ SpringerBriefs in Physics\ (\bibinfo
  {publisher} {Springer International Publishing},\ \bibinfo {year} {2015})\
  \Eprint {https://arxiv.org/abs/arXiv:1210.3781} {arXiv:1210.3781}
  \BibitemShut {NoStop}%
\bibitem [{\citenamefont {{Parisen Toldin}}\ \emph {et~al.}(2009)\citenamefont
  {{Parisen Toldin}}, \citenamefont {{Pelissetto}},\ and\ \citenamefont
  {{Vicari}}}]{PTPV-09}%
  \BibitemOpen
  \bibfield  {author} {\bibinfo {author} {\bibfnamefont {F.}~\bibnamefont
  {{Parisen Toldin}}}, \bibinfo {author} {\bibfnamefont {A.}~\bibnamefont
  {{Pelissetto}}},\ and\ \bibinfo {author} {\bibfnamefont {E.}~\bibnamefont
  {{Vicari}}},\ }\bibfield  {title} {\bibinfo {title} {{Strong-Disorder
  Paramagnetic-Ferromagnetic Fixed Point in the Square-Lattice $\pm J$ Ising
  Model}},\ }\href {https://doi.org/10.1007/s10955-009-9705-5} {\bibfield
  {journal} {\bibinfo  {journal} {\JSF}\ }\textbf {\bibinfo {volume} {135}},\
  \bibinfo {pages} {1039} (\bibinfo {year} {2009})},\ \Eprint
  {https://arxiv.org/abs/0811.2101} {arXiv:0811.2101 [cond-mat.dis-nn]}
  \BibitemShut {NoStop}%
\bibitem [{\citenamefont {{Aharony}}\ and\ \citenamefont
  {{Fisher}}(1983)}]{AF-83}%
  \BibitemOpen
  \bibfield  {author} {\bibinfo {author} {\bibfnamefont {A.}~\bibnamefont
  {{Aharony}}}\ and\ \bibinfo {author} {\bibfnamefont {M.~E.}\ \bibnamefont
  {{Fisher}}},\ }\bibfield  {title} {\bibinfo {title} {{Nonlinear scaling
  fields and corrections to scaling near criticality}},\ }\href
  {https://doi.org/10.1103/PhysRevB.27.4394} {\bibfield  {journal} {\bibinfo
  {journal} {\prb}\ }\textbf {\bibinfo {volume} {27}},\ \bibinfo {pages} {4394}
  (\bibinfo {year} {1983})}\BibitemShut {NoStop}%
\bibitem [{\citenamefont {{Capehart}}\ and\ \citenamefont
  {{Fisher}}(1976)}]{CF-76}%
  \BibitemOpen
  \bibfield  {author} {\bibinfo {author} {\bibfnamefont {T.~W.}\ \bibnamefont
  {{Capehart}}}\ and\ \bibinfo {author} {\bibfnamefont {M.~E.}\ \bibnamefont
  {{Fisher}}},\ }\bibfield  {title} {\bibinfo {title} {{Susceptibility scaling
  functions for ferromagnetic Ising films}},\ }\href
  {https://doi.org/10.1103/PhysRevB.13.5021} {\bibfield  {journal} {\bibinfo
  {journal} {\prb}\ }\textbf {\bibinfo {volume} {13}},\ \bibinfo {pages} {5021}
  (\bibinfo {year} {1976})}\BibitemShut {NoStop}%
\bibitem [{\citenamefont {{Campostrini}}\ \emph {et~al.}(2014)\citenamefont
  {{Campostrini}}, \citenamefont {{Pelissetto}},\ and\ \citenamefont
  {{Vicari}}}]{CPV-14}%
  \BibitemOpen
  \bibfield  {author} {\bibinfo {author} {\bibfnamefont {M.}~\bibnamefont
  {{Campostrini}}}, \bibinfo {author} {\bibfnamefont {A.}~\bibnamefont
  {{Pelissetto}}},\ and\ \bibinfo {author} {\bibfnamefont {E.}~\bibnamefont
  {{Vicari}}},\ }\bibfield  {title} {\bibinfo {title} {{Finite-size scaling at
  quantum transitions}},\ }\href {https://doi.org/10.1103/PhysRevB.89.094516}
  {\bibfield  {journal} {\bibinfo  {journal} {\prb}\ }\textbf {\bibinfo
  {volume} {89}},\ \bibinfo {eid} {094516} (\bibinfo {year} {2014})},\ \Eprint
  {https://arxiv.org/abs/1401.0788} {arXiv:1401.0788 [cond-mat.stat-mech]}
  \BibitemShut {NoStop}%
\bibitem [{\citenamefont {Hasenbusch}(1999)}]{Hasenbusch-99}%
  \BibitemOpen
  \bibfield  {author} {\bibinfo {author} {\bibfnamefont {M.}~\bibnamefont
  {Hasenbusch}},\ }\bibfield  {title} {\bibinfo {title} {{A Monte Carlo study
  of leading order scaling corrections of $\phi^4$ theory on a
  three-dimensional lattice}},\ }\href
  {https://doi.org/10.1088/0305-4470/32/26/304} {\bibfield  {journal} {\bibinfo
   {journal} {\JPAOLD}\ }\textbf {\bibinfo {volume} {32}},\ \bibinfo {pages}
  {4851} (\bibinfo {year} {1999})},\ \Eprint
  {https://arxiv.org/abs/hep-lat/9902026} {hep-lat/9902026} \BibitemShut
  {NoStop}%
\bibitem [{\citenamefont {{Parisen Toldin}}(2011)}]{PT-11}%
  \BibitemOpen
  \bibfield  {author} {\bibinfo {author} {\bibfnamefont {F.}~\bibnamefont
  {{Parisen Toldin}}},\ }\bibfield  {title} {\bibinfo {title} {{Improvement of
  Monte Carlo estimates with covariance-optimized finite-size scaling at fixed
  phenomenological coupling}},\ }\href
  {https://doi.org/10.1103/PhysRevE.84.025703} {\bibfield  {journal} {\bibinfo
  {journal} {\pre}\ }\textbf {\bibinfo {volume} {84}},\ \bibinfo {eid}
  {025703(R)} (\bibinfo {year} {2011})},\ \Eprint
  {https://arxiv.org/abs/1104.2500} {arXiv:1104.2500 [cond-mat.stat-mech]}
  \BibitemShut {NoStop}%
\bibitem [{\citenamefont {{Hasenbusch}}\ \emph {et~al.}(2007)\citenamefont
  {{Hasenbusch}}, \citenamefont {{Parisen Toldin}}, \citenamefont
  {{Pelissetto}},\ and\ \citenamefont {{Vicari}}}]{HPTPV-07}%
  \BibitemOpen
  \bibfield  {author} {\bibinfo {author} {\bibfnamefont {M.}~\bibnamefont
  {{Hasenbusch}}}, \bibinfo {author} {\bibfnamefont {F.}~\bibnamefont {{Parisen
  Toldin}}}, \bibinfo {author} {\bibfnamefont {A.}~\bibnamefont
  {{Pelissetto}}},\ and\ \bibinfo {author} {\bibfnamefont {E.}~\bibnamefont
  {{Vicari}}},\ }\bibfield  {title} {\bibinfo {title} {{The universality class
  of 3D site-diluted and bond-diluted Ising systems}},\ }\href
  {https://doi.org/10.1088/1742-5468/2007/02/P02016} {\bibfield  {journal}
  {\bibinfo  {journal} {\JSTAT}\ } (\bibinfo {year} 2007) {\bibinfo {volume} {P02016}},\
  \bibinfo {pages} {{}} } \Eprint
  {https://arxiv.org/abs/cond-mat/0611707} {cond-mat/0611707} \BibitemShut
  {NoStop}%
\bibitem [{\citenamefont {{Hasenbusch}}(2010{\natexlab{b}})}]{Hasenbusch-10}%
  \BibitemOpen
  \bibfield  {author} {\bibinfo {author} {\bibfnamefont {M.}~\bibnamefont
  {{Hasenbusch}}},\ }\bibfield  {title} {\bibinfo {title} {{Finite size scaling
  study of lattice models in the three-dimensional Ising universality class}},\
  }\href {https://doi.org/10.1103/PhysRevB.82.174433} {\bibfield  {journal}
  {\bibinfo  {journal} {\prb}\ }\textbf {\bibinfo {volume} {82}},\ \bibinfo
  {eid} {174433} (\bibinfo {year} {2010}{\natexlab{b}})},\ \Eprint
  {https://arxiv.org/abs/1004.4486} {arXiv:1004.4486 [cond-mat.stat-mech]}
  \BibitemShut {NoStop}%
\bibitem [{\citenamefont {Hasenbusch}(2019)}]{Hasenbusch-19}%
  \BibitemOpen
  \bibfield  {author} {\bibinfo {author} {\bibfnamefont {M.}~\bibnamefont
  {Hasenbusch}},\ }\bibfield  {title} {\bibinfo {title} {{Monte Carlo study of
  an improved clock model in three dimensions}},\ }\href
  {https://doi.org/10.1103/PhysRevB.100.224517} {\bibfield  {journal} {\bibinfo
   {journal} {\prb}\ }\textbf {\bibinfo {volume} {100}},\ \bibinfo {eid}
  {224517} (\bibinfo {year} {2019})},\ \Eprint
  {https://arxiv.org/abs/1910.05916} {arXiv:1910.05916 [cond-mat.stat-mech]}
  \BibitemShut {NoStop}%
\bibitem [{\citenamefont {Hasenbusch}(1993)}]{Hasenbusch-93}%
  \BibitemOpen
  \bibfield  {author} {\bibinfo {author} {\bibfnamefont {M.}~\bibnamefont
  {Hasenbusch}},\ }\bibfield  {title} {\bibinfo {title} {{Monte Carlo
  simulation with fluctuating boundary conditions}},\ }\href
  {https://doi.org/10.1016/0378-4371(93)90593-S} {\bibfield  {journal}
  {\bibinfo  {journal} {\PAA}\ }\textbf {\bibinfo {volume} {197}},\ \bibinfo
  {pages} {423} (\bibinfo {year} {1993})}\BibitemShut {NoStop}%
\bibitem [{\citenamefont {Campostrini}\ \emph {et~al.}(2001)\citenamefont
  {Campostrini}, \citenamefont {Hasenbusch}, \citenamefont {Pelissetto},
  \citenamefont {Rossi},\ and\ \citenamefont {Vicari}}]{CHPRV-01}%
  \BibitemOpen
  \bibfield  {author} {\bibinfo {author} {\bibfnamefont {M.}~\bibnamefont
  {Campostrini}}, \bibinfo {author} {\bibfnamefont {M.}~\bibnamefont
  {Hasenbusch}}, \bibinfo {author} {\bibfnamefont {A.}~\bibnamefont
  {Pelissetto}}, \bibinfo {author} {\bibfnamefont {P.}~\bibnamefont {Rossi}},\
  and\ \bibinfo {author} {\bibfnamefont {E.}~\bibnamefont {Vicari}},\
  }\bibfield  {title} {\bibinfo {title} {Critical behavior of the
  three-dimensional xy universality class},\ }\href
  {https://doi.org/10.1103/PhysRevB.63.214503} {\bibfield  {journal} {\bibinfo
  {journal} {\prb}\ }\textbf {\bibinfo {volume} {63}},\ \bibinfo {eid} {214503}
  (\bibinfo {year} {2001})},\ \Eprint {https://arxiv.org/abs/cond-mat/0010360}
  {arXiv:cond-mat/0010360 [cond-mat.stat-mech]} \BibitemShut {NoStop}%
\bibitem [{Note1()}]{Note1}%
  \BibitemOpen
  \bibinfo {note} {See Appendix A of Ref.~\cite {PTHAH-14} for a discussion on
  the definition of $\xi $ for a finite lattice, as well as the SM \cite
  {SM}.}\BibitemShut {Stop}%
\bibitem [{\citenamefont {Fisher}\ \emph {et~al.}(1973)\citenamefont {Fisher},
  \citenamefont {Barber},\ and\ \citenamefont {Jasnow}}]{FBJ-73}%
  \BibitemOpen
  \bibfield  {author} {\bibinfo {author} {\bibfnamefont {M.~E.}\ \bibnamefont
  {Fisher}}, \bibinfo {author} {\bibfnamefont {M.~N.}\ \bibnamefont {Barber}},\
  and\ \bibinfo {author} {\bibfnamefont {D.}~\bibnamefont {Jasnow}},\
  }\bibfield  {title} {\bibinfo {title} {Helicity modulus, superfluidity, and
  scaling in isotropic systems},\ }\href
  {https://doi.org/10.1103/PhysRevA.8.1111} {\bibfield  {journal} {\bibinfo
  {journal} {\pra}\ }\textbf {\bibinfo {volume} {8}},\ \bibinfo {pages} {1111}
  (\bibinfo {year} {1973})}\BibitemShut {NoStop}%
\bibitem [{Met()}]{Metlitski-helicity}%
  \BibitemOpen
  \href@noop {} {}\bibinfo {note} {We thank M. Metlitski for pointing out a
  mistake in the formula for $\Upsilon$.}\BibitemShut {Stop}%
\bibitem [{M. Metlitski, private communication.()}]{Metlitski-private}%
  \BibitemOpen
  M. Metlitski, private communication\ \href@noop {} {}\BibitemShut {NoStop}%
\bibitem [{\citenamefont {Parisen~Toldin}\ \emph {et~al.}(2025)\citenamefont
  {Parisen~Toldin}, \citenamefont {Krishnan},\ and\ \citenamefont
  {Metlitski}}]{PTKM-24}%
  \BibitemOpen
  \bibfield  {author} {\bibinfo {author} {\bibfnamefont {F.}~\bibnamefont
  {Parisen~Toldin}}, \bibinfo {author} {\bibfnamefont {A.}~\bibnamefont
  {Krishnan}},\ and\ \bibinfo {author} {\bibfnamefont {M.~A.}\ \bibnamefont
  {Metlitski}},\ }\bibfield  {title} {\bibinfo {title} {{Universal finite-size
  scaling in the extraordinary-log boundary phase of three-dimensional $O(N)$
  model}},\ }\href {https://doi.org/10.1103/PhysRevResearch.7.023052}
  {\bibfield  {journal} {\bibinfo  {journal} {Phys. Rev. Res.}\ }\textbf
  {\bibinfo {volume} {7}},\ \bibinfo {pages} {023052} (\bibinfo {year}
  {2025})},\ \Eprint {https://arxiv.org/abs/2411.05089} {arXiv:2411.05089
  [cond-mat.stat-mech]} \BibitemShut {NoStop}%
\bibitem [{err()}]{erratum}%
  \BibitemOpen
  \href@noop {} {}\bibinfo {note} {A previous version of this Letter was
  missing the factor $2/3$ in front of the $\ln L$ term in the scaling Ansatz
  for $L\Upsilon$. Using the correct scaling form \cite{PTKM-24}, the fitted
  value of $\alpha$ quoted here is in good agreement with estimates obtained
  from other observables. Conclusions and other results of the Letter remain
  unchanged. A detailed investigation of the universal finite-size scaling in
  the extraordinary-log phase is presented in Ref.~\cite{PTKM-24}.}\BibitemShut
  {Stop}%
\bibitem [{\citenamefont {{Hasenbusch}}\ \emph {et~al.}(2008)\citenamefont
  {{Hasenbusch}}, \citenamefont {{Parisen Toldin}}, \citenamefont
  {{Pelissetto}},\ and\ \citenamefont {{Vicari}}}]{HPTPV-08b}%
  \BibitemOpen
  \bibfield  {author} {\bibinfo {author} {\bibfnamefont {M.}~\bibnamefont
  {{Hasenbusch}}}, \bibinfo {author} {\bibfnamefont {F.}~\bibnamefont {{Parisen
  Toldin}}}, \bibinfo {author} {\bibfnamefont {A.}~\bibnamefont
  {{Pelissetto}}},\ and\ \bibinfo {author} {\bibfnamefont {E.}~\bibnamefont
  {{Vicari}}},\ }\bibfield  {title} {\bibinfo {title} {{Universal dependence on
  disorder of two-dimensional randomly diluted and random-bond $\pm J$ Ising
  models}},\ }\href {https://doi.org/10.1103/PhysRevE.78.011110} {\bibfield
  {journal} {\bibinfo  {journal} {\pre}\ }\textbf {\bibinfo {volume} {78}},\
  \bibinfo {eid} {011110} (\bibinfo {year} {2008})},\ \Eprint
  {https://arxiv.org/abs/0804.2788} {arXiv:0804.2788 [cond-mat.dis-nn]}
  \BibitemShut {NoStop}%
\bibitem [{\citenamefont {Diehl}\ and\ \citenamefont
  {Dietrich}(1981)}]{DD-81c}%
  \BibitemOpen
  \bibfield  {author} {\bibinfo {author} {\bibfnamefont {H.~W.}\ \bibnamefont
  {Diehl}}\ and\ \bibinfo {author} {\bibfnamefont {S.}~\bibnamefont
  {Dietrich}},\ }\bibfield  {title} {\bibinfo {title} {Field-theoretical
  approach to multicritical behavior near free surfaces},\ }\href
  {https://doi.org/10.1103/PhysRevB.24.2878} {\bibfield  {journal} {\bibinfo
  {journal} {\prb}\ }\textbf {\bibinfo {volume} {24}},\ \bibinfo {pages} {2878}
  (\bibinfo {year} {1981})}\BibitemShut {NoStop}%
\bibitem [{\citenamefont {Chetyrkin}\ \emph {et~al.}(1983)\citenamefont
  {Chetyrkin}, \citenamefont {Gorishny}, \citenamefont {Larin},\ and\
  \citenamefont {Tkachov}}]{CGLT-83}%
  \BibitemOpen
  \bibfield  {author} {\bibinfo {author} {\bibfnamefont {K.~G.}\ \bibnamefont
  {Chetyrkin}}, \bibinfo {author} {\bibfnamefont {S.~G.}\ \bibnamefont
  {Gorishny}}, \bibinfo {author} {\bibfnamefont {S.~A.}\ \bibnamefont
  {Larin}},\ and\ \bibinfo {author} {\bibfnamefont {F.~V.}\ \bibnamefont
  {Tkachov}},\ }\bibfield  {title} {\bibinfo {title} {Five-loop renormalization
  group calculations in the g{\ensuremath{\phi}} $^{4}$ theory},\ }\href
  {https://doi.org/10.1016/0370-2693(83)90324-6} {\bibfield  {journal}
  {\bibinfo  {journal} {\PLB}\ }\textbf {\bibinfo {volume} {132}},\ \bibinfo
  {pages} {351} (\bibinfo {year} {1983})}\BibitemShut {NoStop}%
\bibitem [{\citenamefont {Kleinert}\ \emph {et~al.}(1991)\citenamefont
  {Kleinert}, \citenamefont {Neu}, \citenamefont {Schulte-Frohlinde},
  \citenamefont {Chetyrkin},\ and\ \citenamefont {Larin}}]{KNSFCL-91}%
  \BibitemOpen
  \bibfield  {author} {\bibinfo {author} {\bibfnamefont {H.}~\bibnamefont
  {Kleinert}}, \bibinfo {author} {\bibfnamefont {J.}~\bibnamefont {Neu}},
  \bibinfo {author} {\bibfnamefont {N.}~\bibnamefont {Schulte-Frohlinde}},
  \bibinfo {author} {\bibfnamefont {K.~G.}\ \bibnamefont {Chetyrkin}},\ and\
  \bibinfo {author} {\bibfnamefont {S.~A.}\ \bibnamefont {Larin}},\ }\bibfield
  {title} {\bibinfo {title} {Five-loop renormalization group functions of
  o(n)-symmetric {\ensuremath{\varphi}}$^{4}$-theory and
  {\ensuremath{\in}}-expansions of critical exponents up to
  {\ensuremath{\in}}$^{5}$},\ }\href
  {https://doi.org/10.1016/0370-2693(91)91009-K} {\bibfield  {journal}
  {\bibinfo  {journal} {\PLB}\ }\textbf {\bibinfo {volume} {272}},\ \bibinfo
  {pages} {39} (\bibinfo {year} {1991})},\ \Eprint
  {https://arxiv.org/abs/hep-th/9503230} {arXiv:hep-th/9503230 [hep-th]}
  \BibitemShut {NoStop}%
\bibitem [{\citenamefont {Kleinert}\ \emph {et~al.}(1993)\citenamefont
  {Kleinert}, \citenamefont {Neu}, \citenamefont {Schulte-Frohlinde},
  \citenamefont {Chetyrkin},\ and\ \citenamefont {Larin}}]{KNSFCL-91_erratum}%
  \BibitemOpen
  \bibfield  {author} {\bibinfo {author} {\bibfnamefont {H.}~\bibnamefont
  {Kleinert}}, \bibinfo {author} {\bibfnamefont {J.}~\bibnamefont {Neu}},
  \bibinfo {author} {\bibfnamefont {N.}~\bibnamefont {Schulte-Frohlinde}},
  \bibinfo {author} {\bibfnamefont {K.~G.}\ \bibnamefont {Chetyrkin}},\ and\
  \bibinfo {author} {\bibfnamefont {S.~A.}\ \bibnamefont {Larin}},\ }\bibfield
  {title} {\bibinfo {title} {{Erratum: Five-loop renormalization group
  functions of O(n)-symmetric {\ensuremath{\varphi}}$^{4}$-theory and
  {\ensuremath{\in}}-expansions of critical exponents up to
  {\ensuremath{\in}}$^{5}$: (Phys. Lett. B 272 (1991) 39)}},\ }\href
  {https://doi.org/10.1016/0370-2693(93)91768-I} {\bibfield  {journal}
  {\bibinfo  {journal} {\PLB}\ }\textbf {\bibinfo {volume} {319}},\ \bibinfo
  {pages} {545} (\bibinfo {year} {1993})},\ \Eprint
  {https://arxiv.org/abs/hep-th/9503230} {arXiv:hep-th/9503230 [hep-th]}
  \BibitemShut {NoStop}%
\bibitem [{\citenamefont {{J\"{u}lich Supercomputing Centre}}(2019)}]{JUWELS}%
  \BibitemOpen
  \bibfield  {author} {\bibinfo {author} {\bibnamefont {{J\"{u}lich
  Supercomputing Centre}}},\ }\bibfield  {title} {\bibinfo {title} {{JUWELS:
  Modular Tier-0/1 Supercomputer at the J\"{u}lich Supercomputing Centre}},\
  }\bibfield  {journal} {\bibinfo  {journal} {Journal of large-scale research
  facilities}\ }\textbf {\bibinfo {volume} {5}},\ \href
  {https://doi.org/10.17815/jlsrf-5-171} {10.17815/jlsrf-5-171} (\bibinfo
  {year} {2019})\BibitemShut {NoStop}%
\bibitem [{\citenamefont {{Parisen Toldin}}\ \emph
  {et~al.}(2015{\natexlab{b}})\citenamefont {{Parisen Toldin}}, \citenamefont
  {{Hohenadler}}, \citenamefont {{Assaad}},\ and\ \citenamefont
  {{Herbut}}}]{PTHAH-14}%
  \BibitemOpen
  \bibfield  {author} {\bibinfo {author} {\bibfnamefont {F.}~\bibnamefont
  {{Parisen Toldin}}}, \bibinfo {author} {\bibfnamefont {M.}~\bibnamefont
  {{Hohenadler}}}, \bibinfo {author} {\bibfnamefont {F.~F.}\ \bibnamefont
  {{Assaad}}},\ and\ \bibinfo {author} {\bibfnamefont {I.~F.}\ \bibnamefont
  {{Herbut}}},\ }\bibfield  {title} {\bibinfo {title} {{Fermionic quantum
  criticality in honeycomb and {$\pi$} -flux Hubbard models: Finite-size
  scaling of renormalization-group-invariant observables from quantum Monte
  Carlo}},\ }\href {https://doi.org/10.1103/PhysRevB.91.165108} {\bibfield
  {journal} {\bibinfo  {journal} {\prb}\ }\textbf {\bibinfo {volume} {91}},\
  \bibinfo {eid} {165108} (\bibinfo {year} {2015}{\natexlab{b}})},\ \Eprint
  {https://arxiv.org/abs/1411.2502} {arXiv:1411.2502 [cond-mat.str-el]}
  \BibitemShut {NoStop}%
\end{thebibliography}%

\clearpage


\onecolumngrid
  \parbox[c][3em][t]{\textwidth}{\centering \large\bf Supplemental Material}
\smallskip
\twocolumngrid

\setcounter{equation}{0}
\renewcommand{\theHequation}{S.\arabic{equation}}
\renewcommand{\theequation}{S.\arabic{equation}}

\setcounter{figure}{0}
\renewcommand{\theHfigure}{S.\arabic{figure}}
\renewcommand{\thefigure}{S.\arabic{figure}}

\setcounter{table}{0}
\renewcommand{\theHtable}{S.\Roman{table}}
\renewcommand{\thetable}{S.\Roman{table}}

\section{Monte Carlo simulations}
We report here some technical details on the MC simulations.
Each elementary update step consists in:
\begin{itemize}
\item a Metropolis sweep over the entire lattice;
\item an overrelaxation sweep;
\item $L$ Wolff single-cluster updates.
\end{itemize}
For the Metropolis step, we update each lattice site in a lexicographic order, and for each site we consider a proposal to update the $\alpha-$component $\phi_i^{(\alpha)}$ of the field $\vec{\phi}_i$ as
\begin{equation}
  \phi_i^{(\alpha)}\rightarrow \phi_i^{(\alpha)} + r\Delta,
  \label{metropolis}
\end{equation}
where $r\in [-1/2,1/2[$ is a uniformly distributed random number and $\Delta$ is chosen to have an good acceptance. We fix $\Delta=2$, for which we have an acceptance of about $48\%$.
On a given lattice site, we loop over $\alpha$, to update all components of $\vec{\phi}_i$.
A Metropolis sweep is followed by an overrelaxation sweep over the entire lattice, where each $\vec{\phi}_i$ is updated as
\begin{equation}
  \vec{\phi}_i \rightarrow 2 \frac{\vec{\phi}_i\cdot\vec{\phi}_{\rm nn}}{\vec{\phi}_{\rm nn}\cdot\vec{\phi}_{\rm nn}}\vec{\phi}_{\rm nn}-\vec{\phi}_i,
  \label{overrelaxation}
\end{equation}
where $\vec{\phi}_{\rm nn}$ is the sum of $\{\vec{\phi}_j\}$ which are nearest neighbor of $i$. The update of Eq.~(\ref{overrelaxation}) is a reflection of $\vec{\phi}_i$ and it is in principle always accepted, since $\vec{\phi}_i\cdot\vec{\phi}_j$ remains unchanged. However, for a small denominator on the right-hand side of Eq.~(\ref{overrelaxation}), such an update is potentially numerically unstable. To fix this, we accept the move only if the variation of $\vec{\phi}_i\cdot\vec{\phi}_j$ does not exceed a threshold, set to $10^{-12}$.
For each Wolff single-cluster update \cite{Wolff-89}, we flip the $\alpha$ component of the fields in a cluster built around a randomly chosen root site, iterating over all components of $\vec{\phi}$.
MC results have been averaged over independent simulations parallelized with the standard Message Passing Interface (MPI). Details on the simulations done are reported in Tables \ref{runs_special} and \ref{runs_extra}.
The integrated autocorrelation time $\tau_{\rm int}$ measured on the surface susceptibility, in units of the update steps, is approximately $\tau\simeq 0.5$ for the MC runs at the special transition, i.e., the hybrid algorithm effectively decorrelates the MC configurations. A larger autocorrelation is instead found in the extraordinary phase, where it grows from $\tau\simeq 1.3$ for $L=8$, $16$ to $\tau\simeq 12$ for $L=256$ and $\tau\approx 20$ for $L=384$.
These estimates of $\tau_{\rm int}$ have been computed with the ``automatic windowing'' algorithm \cite{CS-86,MS-88,Sokal_lecture}.
Error bars are instead estimated with standard Jackknife techniques \cite{Young_notes}, without an explicit determination of the integrated autocorrelation time.

\begin{table}
  \begin{ruledtabular}
    \begin{tabular}{l@{}.@{}.}
      \multicolumn{1}{c}{$L$} & \multicolumn{1}{c}{${\rm Steps} / 10^3$} & \multicolumn{1}{c}{MPI tasks} \\
      \hline
      $16$  & 1000 &   48 \\
      $32$  & 1000 &   48 \\
      $48$  & 1000 &   48 \\
      $64$  & 1000 &   48 \\
      $96$  &  400 &   96 \\
      $128$ &  200 &  240 \\
      $192$ &   50 &  480 \\
    \end{tabular}
  \end{ruledtabular}
  \caption{Details on the MC simulations at the special transition. Each entry corresponds to a single data point in Fig.~\ref{fig.U4}.}
  \label{runs_special}
\end{table}
\begin{table}
  \begin{ruledtabular}
    \begin{tabular}{l..}
      \multicolumn{1}{c}{$L$} & \multicolumn{1}{c}{${\rm Steps} / 10^3$} & \multicolumn{1}{c}{MPI tasks} \\
      \hline
      $8$   & 1000 &   48 \\
      $16$  & 1000 &   48 \\
      $24$  & 1000 &   48 \\
      $32$  &  500 &   48 \\
      $48$  &  100 &   48 \\
      $64$  &  100 &   48 \\
      $96$  &   50 &   48 \\
      $128$ &  100 &   48 \\
      $192$ &   60 &   48 \\
      $256$ &   30 &  192 \\
      $384$ &    5 &  480 \\
    \end{tabular}
  \end{ruledtabular}
  \caption{Details on the MC simulations in the extraordinary phase.}
  \label{runs_extra}
\end{table}

The ratio $Z_a/Z_p$ is computed using the boundary-flip algorithm \cite{Hasenbusch-93}, with the generalization to O(N)-symmetric models discussed in Ref.~\cite{CHPRV-01}.

The second-moment surface correlation length on a finite size $L$ is defined as
\begin{equation}
  \xi = \frac{1}{2\sin(\pi/L)}\sqrt{\frac{\widetilde{C}(0)}{\widetilde{C}(2\pi/L)}-1},
  \label{xi}
\end{equation}
where $\widetilde{C}(p)$ is the Fourier transform of the surface correlations. In Eq.~(\ref{xi}) we average $\widetilde{C}(2\pi/L)$ over the two possible minimum momenta $\vec{p}=(2\pi/L,0)$ and $\vec{p}=(0,2\pi/L)$. We refer to Appendix A of Ref.~\cite{PTHAH-14} for a discussion of the definition of $\xi$ in a finite size.

The helicity modulus $\Upsilon$ describes the response of the system to a twist in the b.c. \cite{FBJ-73}. To fix the notation, we recall that in the model (\ref{model}) we impose periodic BCs on the directions $1$ and $2$, parallel to the surfaces, and open BCs on the remaining direction $3$.
To include a torsion over the components $\alpha$ and $\beta$ of $\vec{\phi}$, we replace
\begin{equation}
  \vec{\phi}_{\vec{x}}\cdot\vec{\phi}_{\vec{x}+\hat{e}_1} \rightarrow \vec{\phi}_{\vec{x}} R_{\alpha,\beta}(\theta)\vec{\phi}_{\vec{x}+\hat{e}_1}, \quad \vec{x}=(x_1=x_{1,f},x_2,x_3),
  \label{torsion}
\end{equation}
where $R_{\alpha,\beta}(\theta)$ is a rotation matrix that rotates the $\alpha$ and $\beta$ components of $\vec{\phi}$ by an angle $\theta$.
In Eq.~(\ref{torsion}) we have slightly generalized the notation, such that $\vec{x}=(x_1,x_2,x_3)$ indicates the lattice site as a three-dimensional vector, and $\hat{e}_1$ is the unit vector in the $1-$direction.
The torsion of Eq.~(\ref{torsion}) results in a $2-$dimensional ``defect'' plane at $x_1=x_{1,f}$ with size $S=L_\parallel L= L^2$. The helicity modulus $\Upsilon$ is defined as \cite{FBJ-73}
\begin{equation}
  \Upsilon \equiv \frac{L}{S} \frac{\partial^2 F(\theta)}{\partial \theta^2}\Big|_{\theta=0}.
  \label{hslicity_def}
\end{equation}
To obtain an easy expression for $\Upsilon$, it is useful, instead of having a plane-defect at $x_1=x_{1,f}$, to smear out the torsion over all length $L$ orthogonal to the plane. Specifically, by a change of variables in the partition sum (a series of rotations), one can write an equivalent Hamiltonian where now the replacement (\ref{torsion}) is
\begin{equation}
  \vec{\phi}_{\vec{x}}\cdot\vec{\phi}_{\vec{x}+\hat{e}_1} \rightarrow \vec{\phi}_{\vec{x}} R_{\alpha,\beta}(\theta/L)\vec{\phi}_{\vec{x}+\hat{e}_1},\qquad \forall \vec{x}.
  \label{torsion_av}
\end{equation}
Using Eq.~(\ref{torsion_av}) the helicity modulus $\Upsilon$ is written as \cite{Metlitski-helicity}
\begin{equation}
  \begin{split}
    \Upsilon &= \frac{1}{L^3}\left( \frac{2}{3}\langle E_1\rangle-\frac{1}{3}\sum_{\alpha < \beta}\Big\langle \left(T^{(\alpha,\beta)}_1\right)^2\Big\rangle\right),\\
    E_1 &= \beta \sum_{\vec{x} \in {\rm bulk}}\vec{\phi}_{\vec{x}}\cdot\vec{\phi}_{\vec{x}+\hat{e}_1}
    + \beta_{s\downarrow}\sum_{\vec{x} \in {\rm surface} \downarrow}\vec{\phi}_{\vec{x}}\cdot\vec{\phi}_{\vec{x}+\hat{e}_1}\\
    &+ \beta_{s\uparrow}\sum_{\vec{x} \in {\rm surface} \uparrow}\vec{\phi}_{\vec{x}}\cdot\vec{\phi}_{\vec{x}+\hat{e}_1}, \\
    T^{(\alpha,\beta)}_1 &= \beta\sum_{\vec{x} \in {\rm bulk}}\left(\phi_{\vec{x}}^{(\alpha)}\phi_{\vec{x}+\hat{e}_1}^{(\beta)} - \phi_{\vec{x}}^{(\beta)}\phi_{\vec{x}+\hat{e}_1}^{(\alpha)}\right)\\
    &+ \beta_{s\downarrow}\sum_{\vec{x} \in {\rm surface} \downarrow}\left(\phi_{\vec{x}}^{(\alpha)}\phi_{\vec{x}+\hat{e}_1}^{(\beta)} - \phi_{\vec{x}}^{(\beta)}\phi_{\vec{x}+\hat{e}_1}^{(\alpha)}\right)\\
    &+ \beta_{s\uparrow}\sum_{\vec{x} \in {\rm surface} \uparrow}\left(\phi_{\vec{x}}^{(\alpha)}\phi_{\vec{x}+\hat{e}_1}^{(\beta)} - \phi_{\vec{x}}^{(\beta)}\phi_{\vec{x}+\hat{e}_1}^{(\alpha)}\right),
  \end{split}
  \label{Upsilon}
\end{equation}
where, to obtain an improved estimator, we have averaged over the $N(N-1)/2=3$ pairs of components $(\alpha, \beta)$ where the torsion is applied.
A further improved estimator of $\Upsilon$ is obtained by averaging over the directions $1$ and $2$ for the torsion:
\begin{equation}
  \begin{split}
    \Upsilon &= \frac{1}{2L^3}\left[\frac{2}{3}\langle E\rangle - \sum_{\hat{e}=\hat{e}_1,\hat{e}_2}\frac{1}{3}\sum_{\alpha<\beta} \langle \left(T^{(\alpha,\beta)}_{\hat{e}}\right)^2\rangle\right],\\
    E &\equiv \beta \sum_{\substack{\vec{x} \in {\rm bulk} \\ \hat{e}=\hat{e}_1,\hat{e}_2}}\vec{\phi}_{\vec{x}}\cdot\vec{\phi}_{\vec{x}+\hat{e}} + \beta_{s\downarrow}\sum_{\substack{\vec{x} \in {\rm surface} \downarrow \\ \hat{e}=\hat{e}_1,\hat{e}_2}}\vec{\phi}_{\vec{x}}\cdot\vec{\phi}_{\vec{x}+\hat{e}}\\
    & + \beta_{s\uparrow}\sum_{\substack{\vec{x} \in {\rm surface}\uparrow \\ \hat{e}=\hat{e}_1,\hat{e}_2}}\vec{\phi}_{\vec{x}}\cdot\vec{\phi}_{\vec{x}+\hat{e}},\\
    T^{(\alpha,\beta)}_{\hat{e}} &\equiv \beta\sum_{\vec{x} \in {\rm bulk}}\left(\phi_{\vec{x}}^{(\alpha)}\phi_{\vec{x}+\hat{e}}^{(\beta)} - \phi_{\vec{x}}^{(\beta)}\phi_{\vec{x}+\hat{e}}^{(\alpha)}\right)\\
    &+ \beta_{s\downarrow}\sum_{\vec{x} \in {\rm surface} \downarrow}\left(\phi_{\vec{x}}^{(\alpha)}\phi_{\vec{x}+\hat{e}}^{(\beta)} - \phi_{\vec{x}}^{(\beta)}\phi_{\vec{x}+\hat{e}}^{(\alpha)}\right)\\
    &+ \beta_{s\uparrow}\sum_{\vec{x} \in {\rm surface} \uparrow}\left(\phi_{\vec{x}}^{(\alpha)}\phi_{\vec{x}+\hat{e}}^{(\beta)} - \phi_{\vec{x}}^{(\beta)}\phi_{\vec{x}+\hat{e}}^{(\alpha)}\right)
  \end{split}
  \label{Upsilon_imp}
\end{equation}
In this work we have used the improved expression of Eq.~(\ref{Upsilon_imp}) and checked that it is consistent with Eq.~(\ref{Upsilon}).

Finally, to validate the program, we have performed a series of tests, the most crucial of which are as follows.  By setting periodic BCs, we have checked the MC results for a small value of $\beta$ with the high-temperature series of Ref.~\cite{CHPRV-02}. Also, for periodic BCs we have reproduced the universal values of the RG-invariants reported in Ref.~\cite{Hasenbusch-20}. Furthermore, we have set $\beta=0$ in the Hamiltonian (\ref{model}) and computed the surface observables which correspond to 2 independent bidimensional surfaces. The results have been successfully compared with MC simulations of the same model in two dimensions, and periodic BCs.

\section{Binder ratio at the special transition}
\begin{figure}[t]
  \centering
  \includegraphics[width=0.8\linewidth]{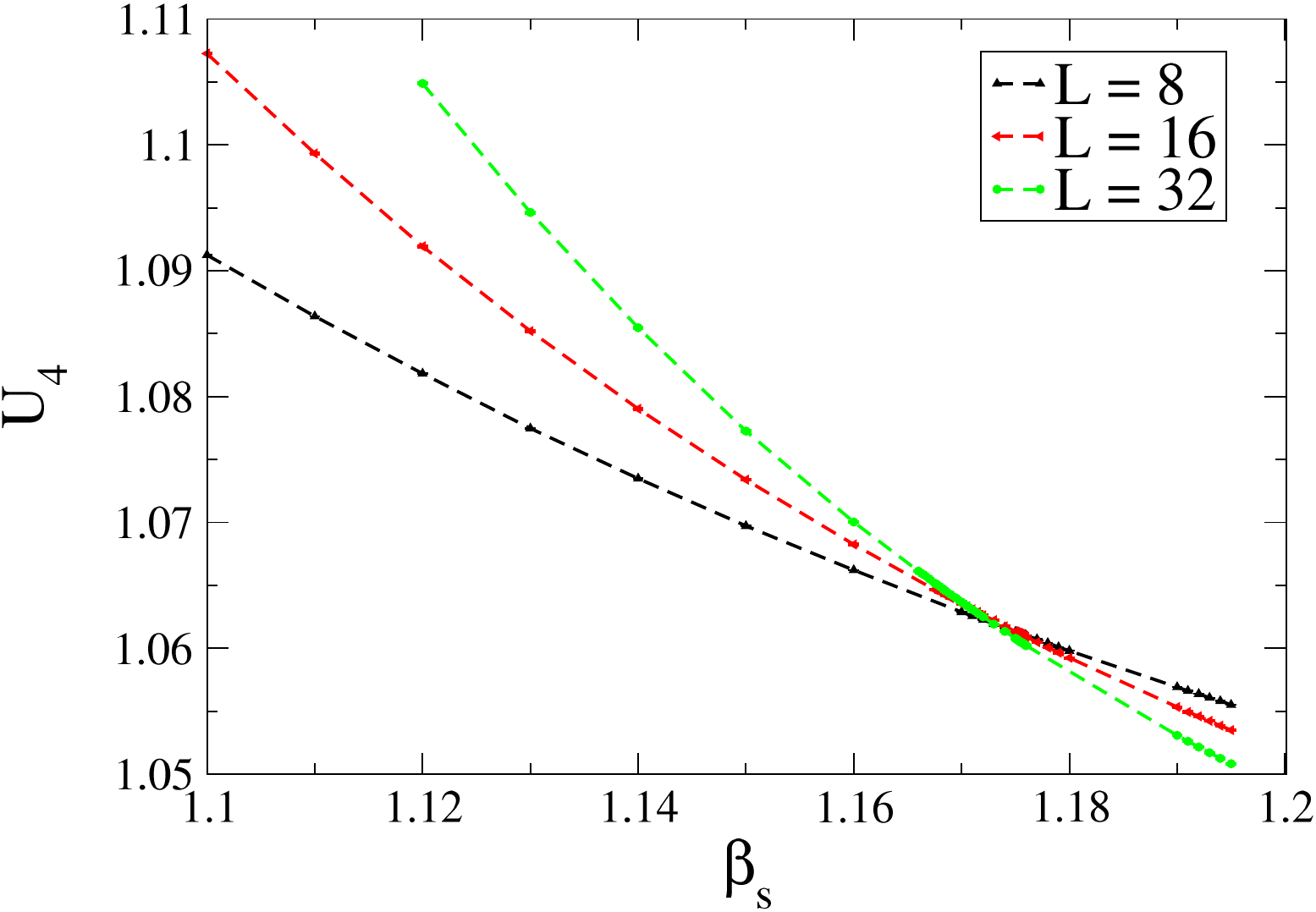}
  \caption{Same as Fig.~\ref{fig.U4} over a larger interval in $\beta_s$ and for lattice sizes $L=8$, $16$, $32$.}
  \label{fig.U4large}
\end{figure}

\begin{table*}
  \begin{ruledtabular}
    \begin{tabular}{l@{}.@{}.@{}.@{}.}
      \multicolumn{1}{c}{$L_{\rm min}$} & \multicolumn{1}{c}{$U_4^*$} & \multicolumn{1}{c}{$\beta_{s,c}$} & \multicolumn{1}{c}{$y_{\rm sp}$} & \multicolumn{1}{c}{$\chi^2/{\rm d.o.f.}$} \\
      \hline
      $16$ &  1.06386(5) & 1.16939(6) & 0.28(2) & 52.1 \\
      $32$ &  1.06463(2) & 1.16847(3) & 0.40(2) & 4.1 \\
      $48$ &  1.06481(3) & 1.16827(3) & 0.41(3) & 1.0 \\
      $64$ &  1.06487(4) & 1.16821(5) & 0.40(4) & 1.1 \\
      \hline
      $16$ &  1.06385(5) & 1.16941(6) & 0.27(2) & 52.1 \\
      $32$ &  1.06463(2) & 1.16847(3) & 0.40(2) & 4.1\\
    \end{tabular}
  \end{ruledtabular}
  \caption{Fits of $U_4$ to Eq.~(\ref{FSS_RGinv_taylor}) at the special transition, with $m=2$, neglecting scaling corrections (above), and with $m=1$ including analytical scaling corrections (below).}
  \label{fits_U4_additional}
\end{table*}
\begin{table*}
  \begin{ruledtabular}
    \begin{tabular}{l@{}.@{}.@{}.@{}.@{}.}
      \multicolumn{1}{c}{$L_{\rm min}$} & \multicolumn{1}{c}{$U_4^*$} & \multicolumn{1}{c}{$\beta_{s,c}$} & \multicolumn{1}{c}{$y_{\rm sp}$} & \multicolumn{1}{c}{$\omega$} & \multicolumn{1}{c}{$\chi^2/{\rm d.o.f.}$} \\
      \hline
      $16$ &  1.0651(2) & 1.1680(1) & 0.40(2) & 1.5(2)   & 0.8 \\
      $32$ &  1.0650(5) & 1.1681(2) & 0.39(2) & 2.4(1.7) & 0.8 \\
    \end{tabular}
  \end{ruledtabular}
  \caption{Fits of $U_4$ to Eq.~(\ref{FSS_RGinv_taylor}) at the special transition, with $m=1$ and a free parameter $\omega$.}
  \label{fits_U4_additional_freeomega}
\end{table*}
In Fig.~\ref{fig.U4large} we show $U_4$ over an interval in $\beta_s$ larger than that of Fig.~\ref{fig.U4}, exhibiting a clear crossing between lattices $L=8$, $16$, and $32$.

In Table \ref{fits_U4_additional} we report additional fits to $U_4$ at the special transition. Fits to the right-hand side of Eq.~(\ref{FSS_RGinv_taylor}) with $m=2$ and neglecting scaling corrections give results almost identical to those of Table \ref{fits_U4}, thereby confirming that $m=1$ is an adequate approximation. To study the influence of analytical scaling corrections, we substitute $(\beta_s-\beta_{s,c})\rightarrow (\beta_s-\beta_{s,c}) + B(\beta_s-\beta_{s,c})^2$ in  Eq.~(\ref{FSS_RGinv_taylor}) and fit the data with $m=1$. The results shown in Table \ref{fits_U4_additional} are identical to those in Table \ref{fits_U4}. Furthermore, for these fits we find that $B=0$ within error bars, hence analytical scaling corrections do not play a role for the MC data considered here. Finally, we consider fits to Eq.~(\ref{FSS_RGinv_taylor}), leaving $\omega$ as a free parameter. Corresponding results shown in Table \ref{fits_U4_additional_freeomega} support a value for $\omega$ compatible with $1$.

\section{Additional analysis at fixed $U_4$ at the special transition}
We consider here the impact on the fitted critical exponents of varying the fixed value of $U_4^* = 1.0652(4)$ between one error bar. In Tables \ref{fits_ysp_additional} and \ref{fits_eta_additional} we report fits for $y_{\rm sp}$ and $\eta_\parallel$, where we fix $U_4=1.0648$ and $U_4=1.0656$.

\begin{table*}
  \begin{ruledtabular}
    \begin{tabular}{ll@{}.@{}.@{}.@{}.}
      & & \multicolumn{2}{c}{$U_4=1.0648$} & \multicolumn{2}{c}{$U_4=1.0656$}\\
      \multicolumn{1}{c}{Obs.} & \multicolumn{1}{c}{$L_{\rm min}$} & \multicolumn{1}{c}{$y_{\rm sp}$} & \multicolumn{1}{c}{$\chi^2/{\rm d.o.f.}$} & \multicolumn{1}{c}{$y_{\rm sp}$} & \multicolumn{1}{c}{$\chi^2/{\rm d.o.f.}$} \\
      \hline
                      & $16$ &  0.3949(7) & 38.5 & 0.3955(7) & 37.0\\
                      & $32$ &  0.380(2)  & 4.7  & 0.381(1)  & 4.6\\
$dU_4/d\beta_s$       & $48$ &  0.373(2)  & 0.2  & 0.374(2)  & 0.2\\
                      & $64$ &  0.372(3)  & 0.2  & 0.373(3)  & 0.2\\
                      & $96$ &  0.369(6)  & 0.03 & 0.370(6)  & 0.03\\[1em]
                      & $16$ &  0.364(3)  & 0.9  & 0.365(3)  & 0.8\\
                      & $32$ &  0.361(5)  & 1.0  & 0.363(5)  & 1.0\\
$d(Z_a/Z_p)/d\beta_s$ & $48$ &  0.363(9)  & 1.3  & 0.365(9)  & 1.3\\
                      & $64$ &  0.34(2)   & 0.02 & 0.35(2)   & 0.02\\
                      & $96$ &  0.34(3)   & 0.02 & 0.35(3)   & 0.05\\
      \hline
                      & $16$ & 0.361(3)  & 0.4  & 0.362(3)   & 0.4\\
$dU_4/d\beta_s$       & $32$ & 0.357(6)  & 0.3  & 0.357(6)   & 0.3\\
                      & $48$ & 0.365(11) & 0.09 & 0.367(11)  & 0.06\\[1em]
                      & $16$ & 0.36(1)   & 1.0  & 0.36(1)    & 1.0\\
$d(Z_a/Z_p)/d\beta_s$ & $32$ & 0.35(2)   & 1.3  & 0.36(2)    & 1.3\\
                      & $48$ & 0.29(4)   & 0.3  & 0.29(4)    & 0.4\\
    \end{tabular}
  \end{ruledtabular}
  \caption{Same as Table \ref{fits_ysp} for fixed $U_4^*=1.0648$ and $U_4^*=1.0656$.}
  \label{fits_ysp_additional}
\end{table*}
\begin{table*}
  \begin{ruledtabular}
    \begin{tabular}{l@{}.@{}.@{}.@{}.}
      & \multicolumn{2}{c}{$U_4=1.0648$} & \multicolumn{2}{c}{$U_4=1.0656$}\\
      \multicolumn{1}{c}{$L_{\rm min}$} & \multicolumn{1}{c}{$\eta_\parallel$} & \multicolumn{1}{c}{$\chi^2/{\rm d.o.f.}$} & \multicolumn{1}{c}{$\eta_\parallel$} & \multicolumn{1}{c}{$\chi^2/{\rm d.o.f.}$} \\
      \hline
      $16$ &   -0.47805(7) & 153.6 & -0.47714(7) & 139.9 \\
      $32$ &   -0.4757(1)  & 13.1  & -0.4749(1)  & 11.4 \\
      $48$ &   -0.4750(2)  & 3.4   & -0.4742(2)  & 2.8 \\
      $64$ &   -0.4746(2)  & 1.5   & -0.4738(3)  & 1.3 \\
      $96$ &   -0.4739(5)  & 0.1   & -0.4733(5)  & 0.2 \\
      \hline
      $16$ &   -0.4725(2)  & 0.4   & -0.4718(2)  & 0.4 \\
      $32$ &   -0.4728(4)  & 0.2   & -0.4721(5)  & 0.2 \\
      $48$ &   -0.4726(8)  & 0.3   & -0.4720(9)  & 0.3 \\
    \end{tabular}
  \end{ruledtabular}
  \caption{Same as Table \ref{fits_eta} for fixed $U_4^*=1.0648$ and $U_4^*=1.0656$.}
  \label{fits_eta_additional}
\end{table*}

\section{Fits of $\beta_{s,c}$}
\begin{table*}
  \begin{ruledtabular}
    \begin{tabular}{l@{}.@{}.@{}.@{}.@{}.@{}.}
      & \multicolumn{2}{c}{$U_4=1.0648$} & \multicolumn{2}{c}{$U_4=1.0652$} & \multicolumn{2}{c}{$U_4=1.0656$}\\
      \multicolumn{1}{c}{$L_{\rm min}$} & \multicolumn{1}{c}{$\beta_{s,c}$} & \multicolumn{1}{c}{$\chi^2/{\rm d.o.f.}$} & \multicolumn{1}{c}{$\beta_{s,c}$} & \multicolumn{1}{c}{$\chi^2/{\rm d.o.f.}$} & \multicolumn{1}{c}{$\beta_{s,c}$} & \multicolumn{1}{c}{$\chi^2/{\rm d.o.f.}$} \\
      \hline
      $16$ & 1.16828(2) & 3.3 & 1.16790(2) & 0.6 & 1.16759(3) & 0.7 \\
      $32$ &            &     & 1.16790(5) & 0.8 & 1.16763(5) & 0.5 \\
    \end{tabular}
  \end{ruledtabular}
  \caption{Fits of the pseudocritical $\beta_{s,c}^{(f)}(L)$ to Eq.~(\ref{FSS_betas_c_fixed}). The fixed-value of $U_4^*=1.0652(4)$ is varied within one error bar.}
  \label{fits_betas_c}
\end{table*}
FSS at fixed RG-invariant $R=R_f$ allows to determine the value of the critical surface coupling $\beta_{s,c}$ at the special transition. For each lattice size $L$, the FSS analysis results in a pseudocritical coupling $\beta_{s,c}^{(f)}(L)$ that converges to $\beta_{s,c}$ for $L\rightarrow\infty$ as
\begin{equation}
  \beta_{s,c}^{(f)}(L) = \beta_{s,c} + AL^{-e},
  \label{FSS_betas_c_fixed}
\end{equation}
where $e=y_{\rm sp}$ for a generic fixed value $R_f$, and $e=y_{\rm sp}+\omega$ if $R_f$ corresponds to the critical one \cite{Hasenbusch-99,PT-11}. In Table \ref{fits_betas_c} we report the results of fit to Eq.~(\ref{FSS_betas_c_fixed}). We consider a variation of $U_4^*=1.0652(4)$ between one error bar. Fits of $\beta_{s,c}^{(f)}(L)$ at the lower bound of $U_4^*$, i.e., at fixed $U_4=1.0648$, deliver a large $\chi^2/d.o.f.$. Furthermore, for $L_{\rm min}=32$ the fit is unstable. For the central value $U_4^*=1.0652$, as well as for the upper bound $U_4^*=1.0656$ fits are overall stable, and with a good $\chi^2/d.o.f.$. Nevertheless, there is a small deviation between the fitted values of $\beta_{s,c}$ at $U_4^*=1.0652$ and at $U_4^*=1.0656$. Therefore, the final estimate of $\beta_{s,c}$ is chosen to be compatible with both these fits.

\section{Fits in the extraordinary phase}
\begin{table*}
  \begin{ruledtabular}
    \begin{tabular}{l@{}.@{}.@{}.@{}.@{}.@{}.}
      \multicolumn{1}{c}{$L_{\rm min}$} & \multicolumn{2}{c}{$\chi_s$} & \multicolumn{2}{c}{$C(L/2)$} & \multicolumn{2}{c}{$C(L/4)$} \\
      & \multicolumn{1}{c}{$q$}  & \multicolumn{1}{c}{$\chi^2/{\rm d.o.f.}$} & \multicolumn{1}{c}{$q$}  & \multicolumn{1}{c}{$\chi^2/{\rm d.o.f.}$} & \multicolumn{1}{c}{$q$} & \multicolumn{1}{c}{$\chi^2/{\rm d.o.f.}$} \\
      \hline
      $8$   & 2.254(4)  &      2.6  & 2.049(4) &     37.7  & 2.036(2) &    161.7 \\
      $16$  & 2.243(6)  &      2.3  & 2.145(7) &      1.9  & 2.149(4) &      8.3 \\
      $24$  & 2.224(10) &      1.7  & 2.16(1)  &      2.0  & 2.188(7) &      1.9 \\
      $32$  & 2.22(2)   &      2.1  & 2.17(2)  &      2.2  & 2.21(2)  &      1.5 \\
      $48$  & 2.18(4)   &      2.1  & 2.12(4)  &      2.3  & 2.19(3)  &      1.6 \\
      $64$  & 2.10(5)   &      1.8  & 2.03(7)  &      2.1  & 2.15(4)  &      1.5 \\
      $96$  & 2.1(2)    &      2.6  & 2.1(2)   &      3.1  & 2.18(8)  &      2.2 \\
      $128$ & 2.6(3)    &      0.3  & 2.7(4)   &      0.1  & 2.4(2)   &      0.2 \\
    \end{tabular}
  \end{ruledtabular}
  \caption{Fits for $q$ in the extraordinary phase, as extracted from the surface susceptibility $\chi_s$ and the surface correlations $C(L/2)$ and $C(L/4)$. $L_{\rm min}$ is the minimum lattice size taken into account.}
  \label{fits_q_extra}
\end{table*}
In Table \ref{fits_q_extra} we report fit results of the surface susceptibility $\chi_s$ to $AL^2\ln(L/l_0)^{-q}$, and of the correlations $C(L/2)$, $C(L/4)$ to $A\ln(L/l_0)^{-q}$, leaving $A$, $l_0$, and $q$ as free parameters; in practice, to obtain fit stability, we employ as Ansatz $AL^2(\ln(L)+c)^{-q}$ and $A(\ln(L)+c)^{-q}$.
$L_{\rm min}$ is the minimum lattice size taken into account.

\begin{table}
  \begin{ruledtabular}
    \begin{tabular}{ll@{}.@{}.}
      \multicolumn{1}{c}{Obs.} & \multicolumn{1}{c}{$L_{\rm min}$} & \multicolumn{1}{c}{$\alpha$} & \multicolumn{1}{c}{$\chi^2/{\rm d.o.f.}$} \\
      \hline
           & $8$   & 0.0795(1) & 4406.0 \\
           & $16$  & 0.1103(2) & 474.6 \\
           & $24$  & 0.1228(3) & 100.8 \\
           & $32$  & 0.1291(5) & 30.0 \\
$(\xi/L)^2$ & $48$  & 0.1355(7) & 7.1 \\
           & $64$  & 0.1384(9) & 2.6 \\
           & $96$  & 0.142(2)  & 0.4 \\
           & $128$ & 0.142(2)  & 0.4 \\
           & $192$ & 0.145(5)  & 0.2 \\
      \hline
             & $8$ & 0.1160(2) & 110.7 \\
             & $16$ & 0.1278(5) & 9.9 \\
             & $24$ & 0.135(1) & 1.8 \\
$\Upsilon L$ & $32$ & 0.136(2) & 1.9 \\
             & $48$ & 0.141(3) & 1.4 \\
             & $64$ & 0.144(5) & 1.5 \\
             & $96$ & 0.156(8) & 0.8 \\
             & $128$& 0.167(12)  & 0.3 \\
    \end{tabular}
  \end{ruledtabular}
  \caption{Fits of $\alpha$ as extracted from $(\xi/L)^2$ and $\Upsilon L$, as a function of the minimum lattice size taken into account.}
  \label{fits_alpha}
\end{table}
In Table \ref{fits_alpha} we report fit results of $(\xi/L)^2$ to $(\alpha/2)\ln(L) + B$ and of $\Upsilon L$ to $2\alpha\ln(L) + B$ \cite{Metlitski-private}, leaving $\alpha$ and $B$ as free parameters.

\section{Comparison with field-theory results}
For the O(N) UC, the two-loops $\varepsilon-$expansion series for $\eta_\parallel$ at the special transition and the crossover exponent $\Phi\equiv y_{\rm sp}\nu$ are \cite{DD-81c,Diehl-86}
\begin{equation}
  \eta_\parallel = -\frac{n+2}{n+8}\varepsilon + \frac{5(4-n)(2+n)}{2(8+n)^3}\varepsilon^2+ O(\varepsilon)^3,
  \label{eta_parallel_eexp}
\end{equation}

\begin{equation}
  \begin{split}
    \Phi = &\frac{1}{2} -\frac{N+2}{4(N+8)}\varepsilon \\
    &+\frac{N+2}{8(N+8)^3}\left[8\pi^2(N+8)-(N^2+35N+156)\right]\varepsilon^2 \\
    &+ O(\varepsilon)^3.
  \end{split}
  \label{Phi_eexp}
\end{equation}
Using the well-known $\varepsilon-$expansion result of $1/\nu$ \cite{CGLT-83,KNSFCL-91,*KNSFCL-91_erratum}
\begin{equation}
  1/\nu = 2-\frac{N+2}{N+8}\varepsilon - \frac{(N+2)(13N+44)}{2(N+8)^3}\varepsilon^2 + O(\varepsilon)^3,
\end{equation}
the $\varepsilon-$expansion series for $y_{\rm sp}$ is
\begin{equation}
  \begin{split}
    y_{\rm sp} &= 1 - \frac{N+2}{N+8}\varepsilon+\frac{(N+2)(32+4N)\pi^2-19N-92}{2(N+8)^3}\varepsilon^2\\
    &+O(\varepsilon)^3.
  \end{split}
  \label{ysp_eexp}
\end{equation}
Setting $N=3$ in Eq.~(\ref{eta_parallel_eexp}) and Eq.~(\ref{ysp_eexp}), a simple summation to $\varepsilon=1$ gives $\eta_\parallel=-0.445$, $y_{\rm sp}= 1.081$. Employing a $[1/1]$ Pad\'e resummation, we find $\eta_\parallel=-0.445$, $y_{\rm sp}= 0.791$. Alternatively, one can analyze the series of $\Phi$, and compute $y_{\rm sp}=\Phi/\nu$ using $\nu=0.7112$ \cite{CHPRV-02}. In this case, we obtain $y_{\rm sp}=0.938$ for a direct summation, and $y_{\rm sp}=0.657$ for a $[1/1]$ Pad\'e approximation.
\end{document}